\documentclass[useAMS,usenatbib]{mn2e}
\usepackage{graphicx}
\usepackage{epstopdf}
\usepackage{amsmath}
\usepackage{amssymb}
\usepackage{longtable}
\usepackage{lscape}
\usepackage{rotating}
\usepackage{natbib}
\usepackage{epsfig}
\usepackage{array}
\newcommand{\lsim}{\lower 2pt \hbox{$\, \buildrel {\scriptstyle<}\over {\scriptstyle \sim}\,$}}  
\newcommand{\gsim}{\lower 2pt\hbox{$\, \buildrel {\scriptstyle >}\over {\scriptstyle \sim}\,$}}

\newcommand{\oxythr}{{[OIII]}}

\title[HST {[OIII]} imaging of nearby ULIRGs]{Quantifying the AGN-driven outflows in ULIRGs (QUADROS) II: evidence for compact outflow regions from HST [OIII] imaging observations}
\author[C. Tadhunter et al.]{C. Tadhunter$^{1}$\thanks{E-mail: c.tadhunter@sheffield.ac.uk}, J. Rodr\'iguez Zaur\'in$^{1}$, M. Rose$^{1}$, R. A. W. Spence$^{1}$, \newauthor D. Batcheldor$^{2}$, M.A. Berg$^{3}$, C. Ramos Almeida$^{4,5}$, H.W.W. Spoon$^{6}$, \newauthor W. Sparks$^{7}$, M. Chiaberge$^{7,8}$ \\
$^1$Department of Physics and Astronomy, University of Sheffield, Sheffield S3
7RH, UK \\
$^2$ Physics and Space Sciences Department, Florida Institute of Technology, 150 West University Boulevard, Melbourne, FL 32901, USA \\
$^3$ Department of Physics, University of Notre Dame, Notre Dame, IN, USA \\
$^4$Instituto de Astrof\' isica de Canarias, Calle V\' ia L\'actea, s/n, E-38205, La Laguna, Tenerife, Spain \\
$^5$Departamento de Astrof\' isica, Universidad de La Laguna, E-38205, La Laguna, Tenerife, Spain \\
$^6$ Cornell Center for Astrophysics and Planetary Science, Space Sciences Building, Ithaca, NY 14853, USA \\
$^7$ Space Telescope Science Institute, 3700 San Martin Drive, Baltimore, MD 21218, USA \\
$^8$ Center for Astrophysical Sciences, Johns Hopkins University, 3400 N. Charles Street, Baltimore, MD 21218, USA}
\begin{document}


\maketitle

\label{firstpage}

\begin{abstract}{The true importance of the warm, AGN-driven outflows for the evolution of galaxies remains uncertain. Measurements of the radial extents of the outflows are key for quantifying their masses and kinetic powers, and also establishing whether the AGN outflows are galaxy-wide.  Therefore, as part of a larger project to investigate the significance of warm, AGN-driven outflows in the most rapidly evolving galaxies in the local universe, here we present deep {\it Hubble Space Telescope} ({\it HST}) narrow-band [OIII]$\lambda$5007 observations of a complete sample of 8 nearby ULIRGs with optical AGN nuclei. Combined with the complementary information provided by our ground-based spectroscopy, the {\it HST} images show that the warm gas outflows are relatively compact for most of the objects in the sample: in three objects the outflow regions are barely resolved at the resolution of {\it HST} ($0.065 < R_{[OIII]} <0.12$\,kpc); in a further four cases the outflows are spatially resolved but with flux weighted mean radii in the range $0.65 < R_{[OIII]} < 1.2$\,kpc; and in only one object (Mrk273) is there clear evidence for a more extended outflow, with a maximum extent of $R_{[OIII]}\sim5$\,kpc. Overall, our observations show little evidence for the galaxy-wide outflows  predicted by some models of AGN feedback. }

\end{abstract}

\begin{keywords}
Galaxies: evolution -- galaxies: starburst -- galaxies:active
\end{keywords}

%

\section{Introduction}

Ultra Luminous Infrared Galaxies (ULIRGs:
$L_{IR} > 10^{12}$\,\,L$\odot$; \citealt{sanders96}) represent the peaks of star formation activity in major, gas-rich mergers, in which both the galaxy bulges
and supermassive black holes are growing rapidly. As such, they are among the most actively evolving systems 
in the local universe. Indeed, they represent just the situation modelled 
in hydrodynamic simulations of the co-evolution of black holes and 
their host galaxies that incorporate the feedback effect associated with AGN-driven outflows (\citealt{dimatteo05}; \citealt{springel05}; 
\citealt{johansson09}). Therefore, the nearby
ULIRGs are ideal objects to test whether AGN outflows have a major impact on
the evolution of galaxies. 

Observations of high-ionization [OIII], [NeV] and [NeIII] emission lines at optical (\citealt{holt03}; \citealt{holt11}; \citealt{rz13}, hereafter RZ13) and mid-IR (\citealt{spoon09a}; \citealt{spoon09b}) wavelengths have revealed the presence of warm outflows in the nuclei of a large proportion (94\%: RZ13) of nearby ULIRGs in which Seyfert-like AGN have been detected at optical wavelengths; such outflows are
rare in ULIRGs that lack optically-detected AGN (RZ13; see also Arribas et al. 2014). 
Moreover, the  measured emission line ratios of the detected outflows are consistent with AGN photoionization rather than shock ionization or photoionization by stars, and the outflows are concentrated in the nuclear regions of the galaxies. Therefore it is highly likely that these outflows are driven by the AGN rather than starbursts. 
However, despite the common presence of the  warm nuclear outflows in ULIRGs with optical AGN nuclei, their true significance for the evolution of the 
host galaxies has yet to be determined. In particular, it has so far proved challenging to quantify the masses, kinetic powers and momenta of the near-nuclear outflows, because their radial scales, densities, reddenings and emission line luminosities have been difficult to measure accurately using the
existing observations \citep[see discussion in][]{rz13,harrison18}. 

Determining the radial scales of the outflows is particularly important, because not only is the radius a key parameter in the equations used to calculate the outflow properties (e.g. RZ13), but measurements of the radius also provide a direct indication of 
the ``sphere of influence'' -- the proportion of the host galaxy currently affected by the outflow. In this
context, we note that analytic models of energy-driven
AGN outflows indicate that they can reach galaxy-wide scales of ($>$10\,kpc) on timescales comparable
with the typical $10^6 - 10^8$\,yr lifetimes of luminous AGN activity \citep{king11}, resulting in the 
removal of the ambient gas from the galaxies' bulges. However, the true radial extends of warm outflows driven by AGN are  a subject of considerable debate in the literature (e.g.
compare the results of \citealt{harrison12}, \citealt{liu13} and \citealt{harrison14}  with those of \citealt{husemann16}, \citealt{karouzos16},\citealt{vm16} and \citealt{fischer18}).
Unfortunately, for most of the low-redshift ULIRGs with optical AGN nuclei the outflows are unresolved in existing ground-based optical and Spitzer mid-IR spectra ($\sim$0.7-2.0 arcsec resolution for the ground-based spectra of RZ13, and $\sim$6 arcsec resolution for the Spitzer spectroscopy of \citealt{spoon09b}). 

Therefore, in order to properly quantify the warm, AGN-driven outflows in ULIRGs, we are
undertaking a programme that combines high resolution {\it Hubble Space Telescope (HST)} observations -- to measure the radii and geometries of the outflows -- and wide-spectral-coverage spectroscopic observations -- to measure their densities, reddenings and emission line luminosities: the Quantifying Ulirg Agn-DRiven OutflowS (QUADROS) project. Our full sample for this project comprises 15 objects and is described in detail in \citet{rose17}, where we also present the results of deep VLT/Xshooter spectroscopic observations for 7 of the sample. In this second paper in the series, we present {\it HST} Advanced Camera for Surveys (ACS)
imaging observations for a complete sample of 8 nearby ULIRGs with optical AGN nuclei, concentrating on the radial scales of the emission line regions, while in Spence et al. (2018; Paper III) we will present
the spectroscopic results for a further 8 objects, and investigate possible links between the warm outflows and the properties of the AGN and host galaxies.

As discussed in sections 3.1 and 3.2, our method involves measuring the radial extents of the near-nuclear emission-line gas directly from our high-resolution {\it HST} images, then using the complementary information provided by our ground-based spectra to estimate the extents of the warm outflows.  We emphasise that our results refer to the radial scales of emission line outflows that often dominate the {\it nuclear} spectra of ULIRGs in ground-based observations, rather than to the more extended and quiescent emission line gas that is sometimes present in the extended narrow-line regions (ENLR) of such objects \citep[e.g.][]{spence16}. Moreover, although there is
no generally agreed kinematic criterion in the literature for what constitutes an outflow, here we take 
any [OIII] emission line component with line-width $FWHM > 500$\,km s$^{-1}$ to represent an outflow. 
Throughout the paper
we adopt a cosmology with H$_{0}$ = 73 km s$^{-1}$, $\Omega_{\rm m}$ = 0.27 and
$\Omega_{\Lambda}$ = 0.73.

\section{Sample, observations and data reduction}

\subsection{The sample}

\begin{table*}
\centering
\small
\begin{tabular}{lccccclcccl}
\hline\hline
Name & & & & & & & & & &\\
IRAS & z & Scale & log(L$_{\rm IR}$) & Filter &[OIII]/ & $\lambda_{c}$ &FWHM &Exp &F[OIII]/10$^{-14}$ &F$_{WHT}$/ \\ 
FSC  &   & kpc/'' &(log(L$_{\odot}$)) & &Cont & (\AA) & (\AA) & sec &(erg s$^{-1}$ cm$^{-2}$) &F$_{HST}$\\ 
(1)  & (2)& (3)      & (4)       & (5) & (6) & (7) & (8) & (9) & (10) & (11)\\
\hline
F13428+5608	&	0.037	&	0.727	&	12.18	&	F550M	&	Cont	&	5581	&	384	&	716	&	12	&	0.24	\\
	&			&	&	&	FR505N	&	[OIII]	&	5176	&	92	&	1356	&	&	\\			
F13443+0802	&	0.135	&	2.320	&	12.23	&	FR647M	&	Cont	&	6133	&	379	&	678	&	0.89	&	0.88$^{*}$	\\
	&			&	&	&	FR551N	&	[OIII]	&	5664	&	90	&	1290	&	&	\\			
F13451+1232	&	0.122	&	2.120	&	12.36	&	FR459M	&	Cont	&	5093	&	345	&	2560	&	7.4	&	0.44	\\
	&			&	&	&	F550M	&	[OIII]	&	5580	&	389	&	2480	&	&	\\			
F14394+5332	&	0.104	&	1.849	&	12.12	&	FR647M	&	Cont	&	5967	&	367	&	680	&	1.8	&	0.87	\\
	&			&	&	&	FR551N	&	[OIII]	&	5510	&	82	&	1356	&	&	\\			
F15130--1958	&	0.109	&	1.917	&	12.17	&	FR647M	&	Cont	&	5590	&	367	&	680	&	2.0	&	0.64	\\
	&			&	&	&	FR551N	&	[OIII]  &	5531	&	83	&	1335		&	&	\\			
F16156+0146	&	0.132	&	2.262	&	12.12	&	FR647M	&	Cont	&	6116	&	378	&	680	&	2.7	&	0.68	\\
	&			&	&	&	FR551N	&	[OIII]	&	5647	&	89	&	1290	&	&	\\			
F17044+6720	&	0.135	&	2.300	&	12.21	&	FR647M	&	Cont	&	6131	&	379	&	720	&	0.55	&	0.75	\\
	&			&	&	&	FR551N	&	[OIII]	&	5662	&	90	&	1440	&	&	\\			
F17179+5444	&	0.147	&	2.471	&	12.28	&	FR647M	&	Cont	&	6196	&	383	&	678	&	1.1	&	0.80	\\
	&			&	&	&	FR551N	&	[OIII]	&	5721	&	95	&	1356	&	&	\\			

\end{tabular}
\caption []{ The sample of 8 ULIRGs discussed in this paper. Col (1): object
  designation in the IRAS Faint Source Catalogue Database (FSC). Col (2):
  optical redshifts from Kim \& Sanders (1998). Col (3): Scale (kpc arcsec$^{-1}$) for our adopted cosmology
  (H$_{0}$ = 73 km s$^{-1}$, $\Omega_{\rm m}$ = 0.27 and $\Omega_{\Lambda}$ =
  0.73). Col (4): IR luminosity from Kim and Sanders (1998) adapted to our
  cosmology. Col (5)-(6): Filter used and the corresponding {\it HST}
  observation. Col (7)-(9): central wavelength, bandwidth and exposure time. Col (10): Galactic reddening corrected [OIII]$\lambda$5007 flux
measured in a 5\,kpc diameter aperture centred on the main nucleus. Col (11): HST/WHT [OIII]$\lambda$5007 flux ratios that compare the fluxes obtained using ground-based spectroscopy by RZ13 with those measured from the
{\it HST} images (see text for details). $*$ Note: RZ13 did not include F13443+0802 in their study, so in this case
we have compared with the total VLT/Xshooter [OIII]$\lambda$5007 for this source, as measured by Rose et al. (2017).}
\label{Sample}
\end{table*}

%
%

The full QUADROS sample consists of 15 nearby ULIRGs with $z < 0.175$ \citep[see][for description]{rose17}. In this paper we describe {\it HST} observations of a complete subset
of the QUADROS sample which comprises all
8 objects from the 1Jy Sample of ULIRGs \citep{kim98} classified as Sy2 on the basis of their optical spectra
\citep{yuan10}, with redshifts z $<$ 0.15, right ascensions 13:40 $<$ RA $<$
23:20~hr and declinations $\delta >$ -25 degrees. The upper redshift limit is
chosen to ensure that there is a reasonable chance of spatially resolving the
outflow regions (0.1 arcseconds corresponds to
$\sim$0.25\,kpc at z = 0.15), while the RA limits were set by scheduling
constraints imposed in {\it HST} cycle during which the observations were taken (Cycle 20). Note that objects with broad-line AGN are excluded from our 
{\it HST} sample because the [OIII] emission lines are potentially contaminated by FeII emission
from the BLR, making it difficult to map the warm outflows in the near-nuclear regions. 
Table \ref{Sample} presents some basic
information for the 8 ULIRGs in our 
{\it HST} sample.

\subsection{HST observations and data reduction}

\begin{figure*}
\centering
\begin{tabular}{ccc}
\hspace{-0.4cm}\includegraphics[width=0.36\textwidth]{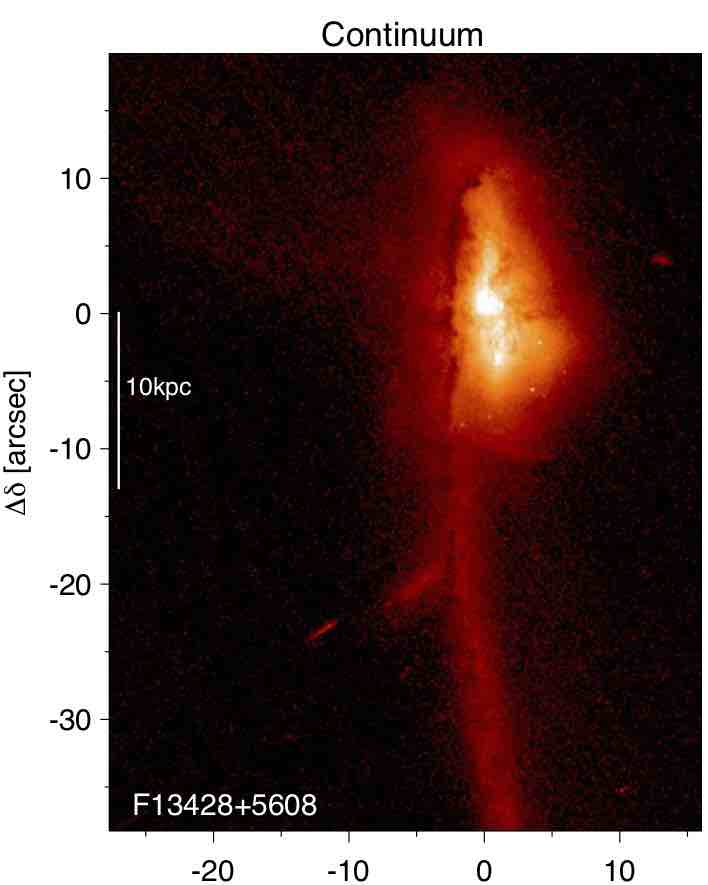}&
\hspace{-1cm}\includegraphics[width=0.31\textwidth]{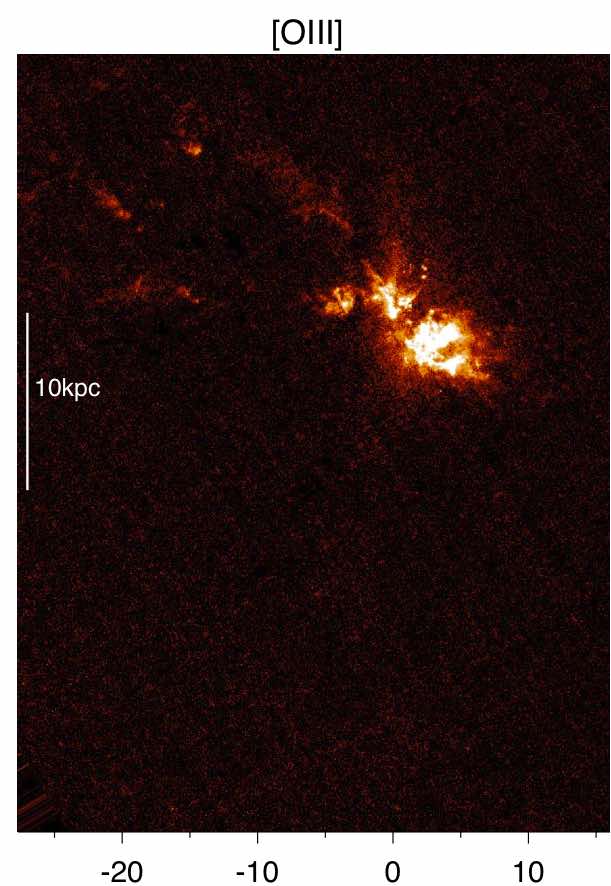}&
\hspace{-0.8cm}\includegraphics[width=0.301\textwidth]{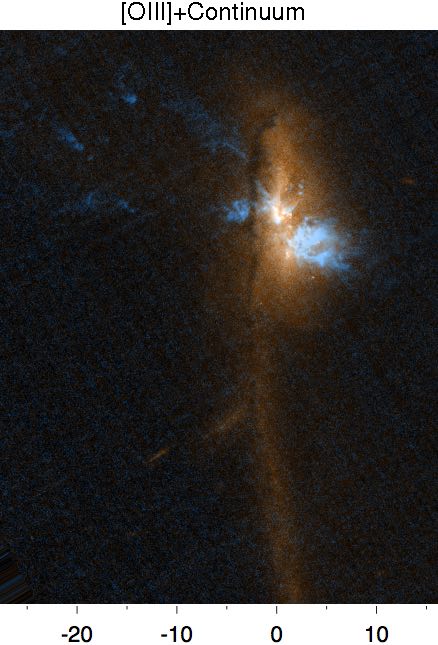}\\
\hspace{-0.3cm}\includegraphics[width=0.37\textwidth]{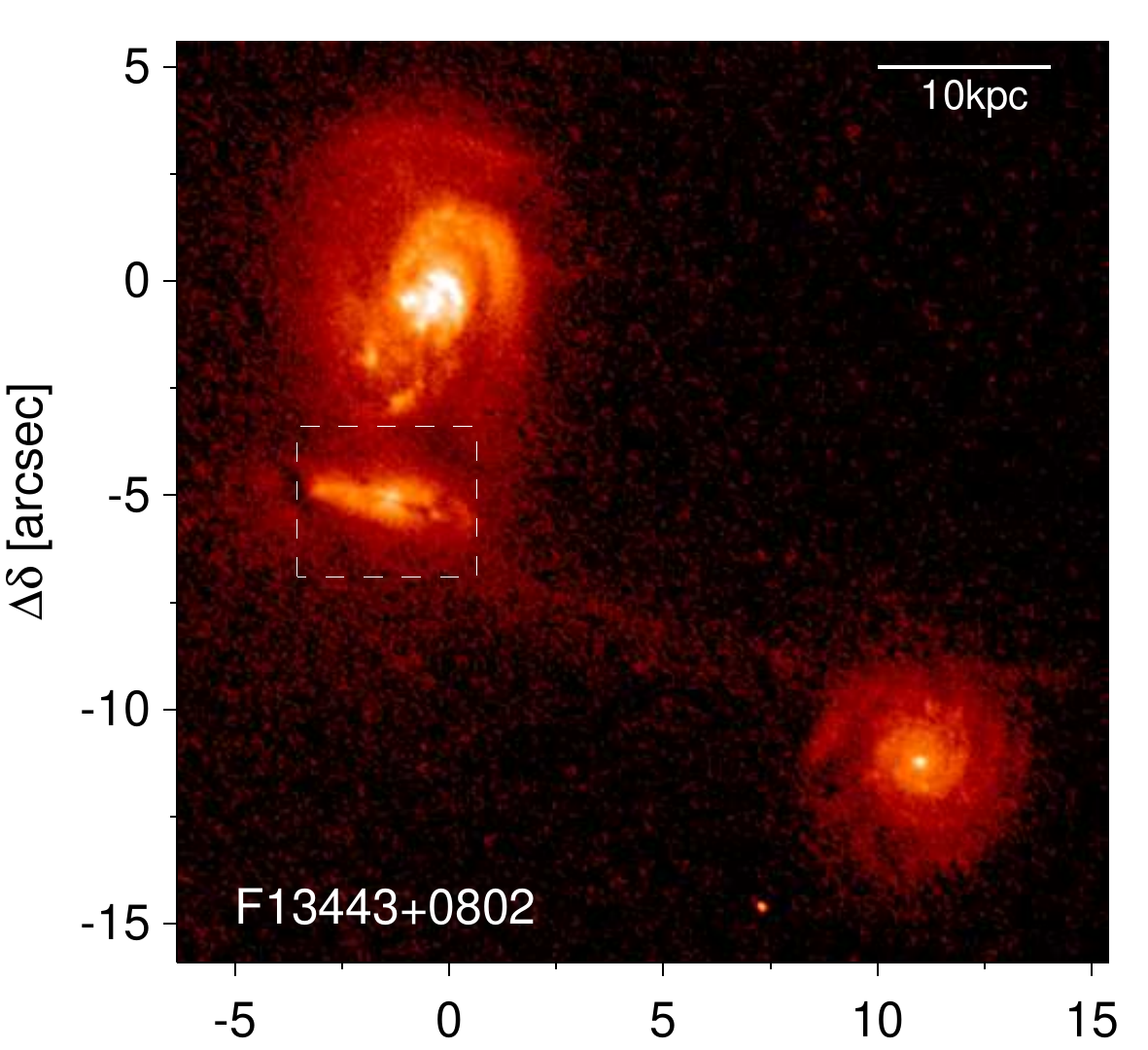}&
\hspace{-0.5cm}\includegraphics[width=0.325\textwidth]{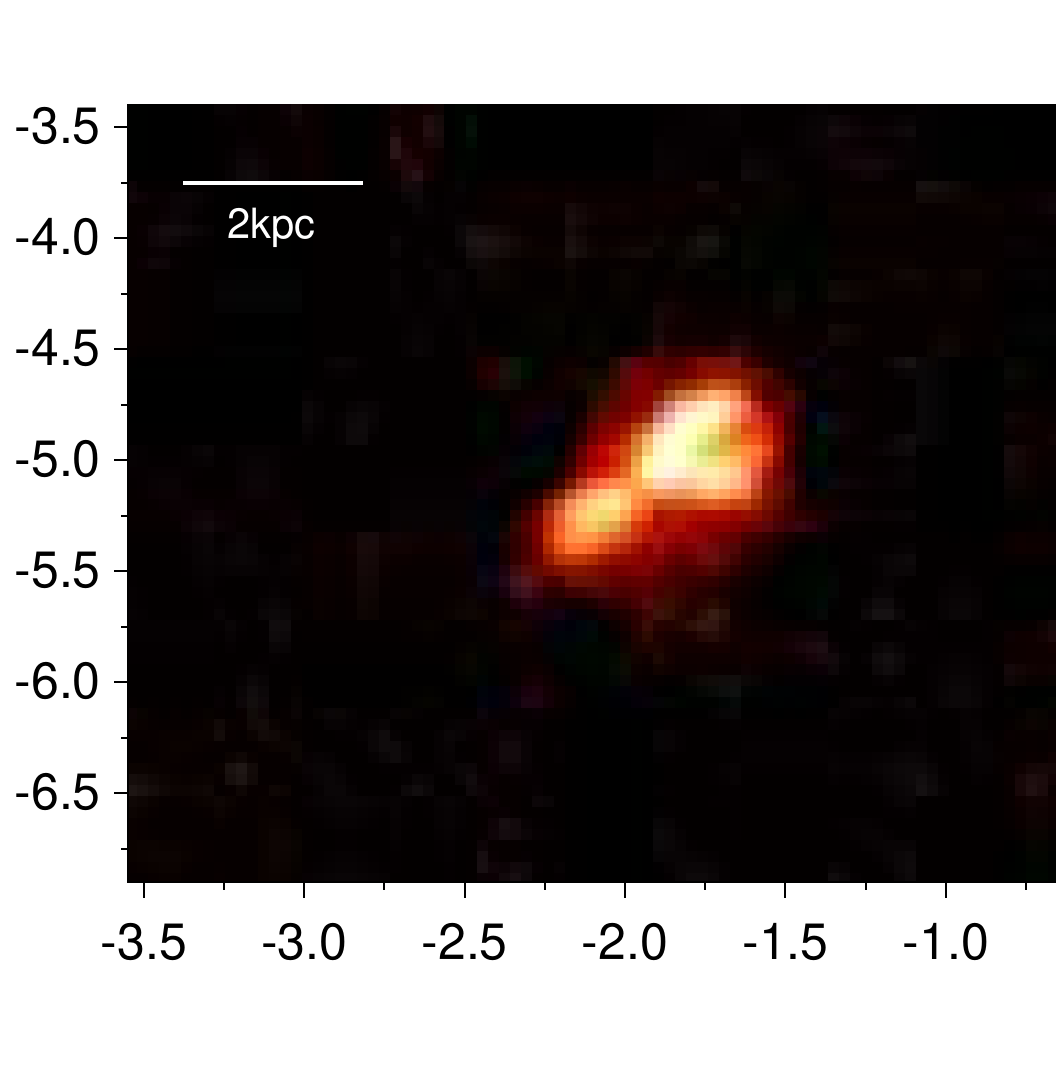}&
\hspace{-0.3cm}\includegraphics[width=0.3\textwidth]{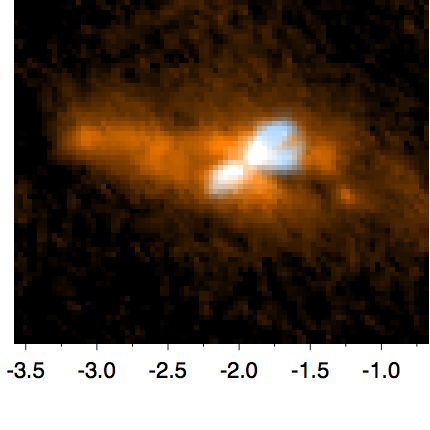}\\
\hspace{-0.4cm}\includegraphics[width=0.37\textwidth]{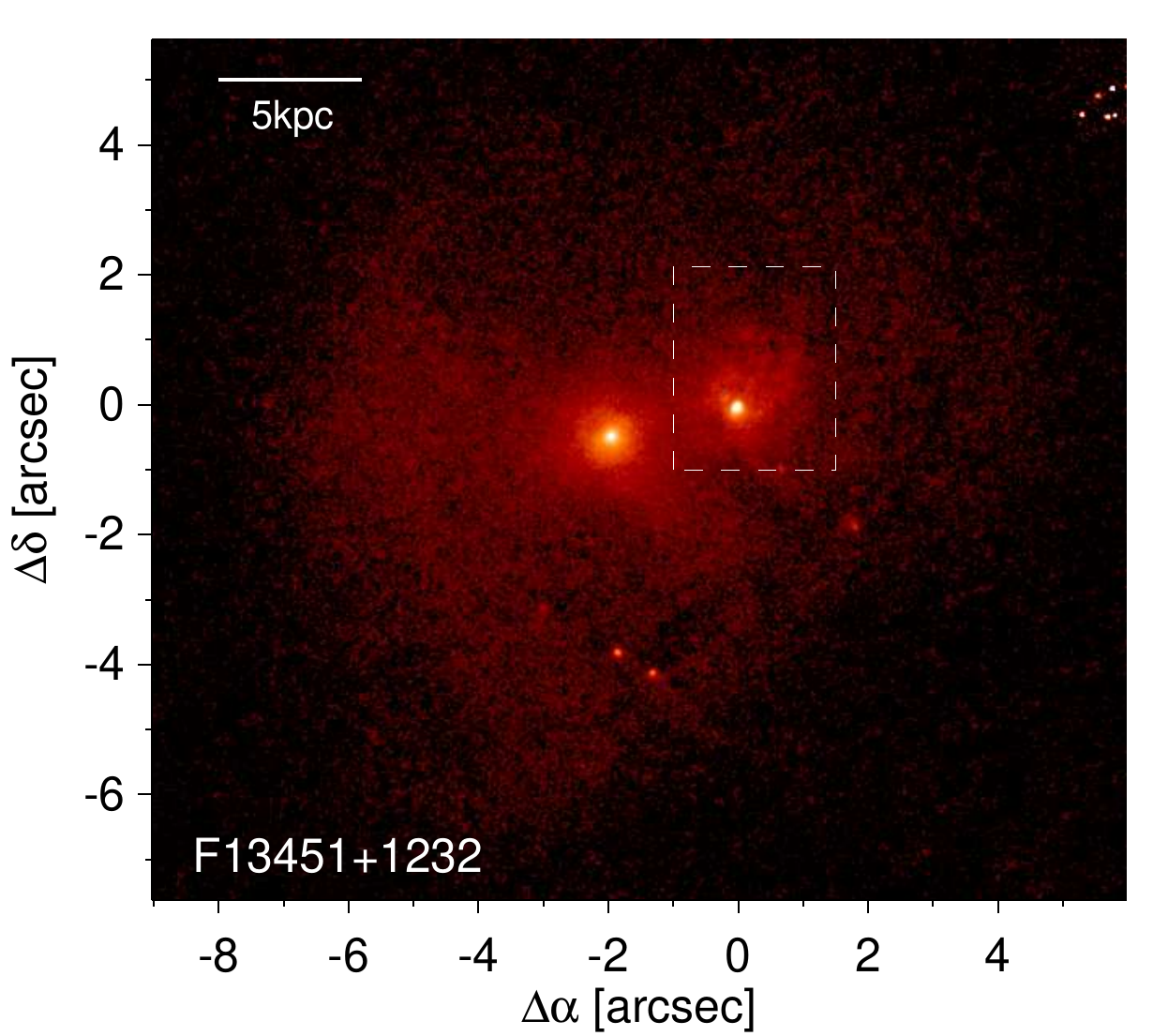}&
\hspace{-0.cm}\includegraphics[width=0.29\textwidth]{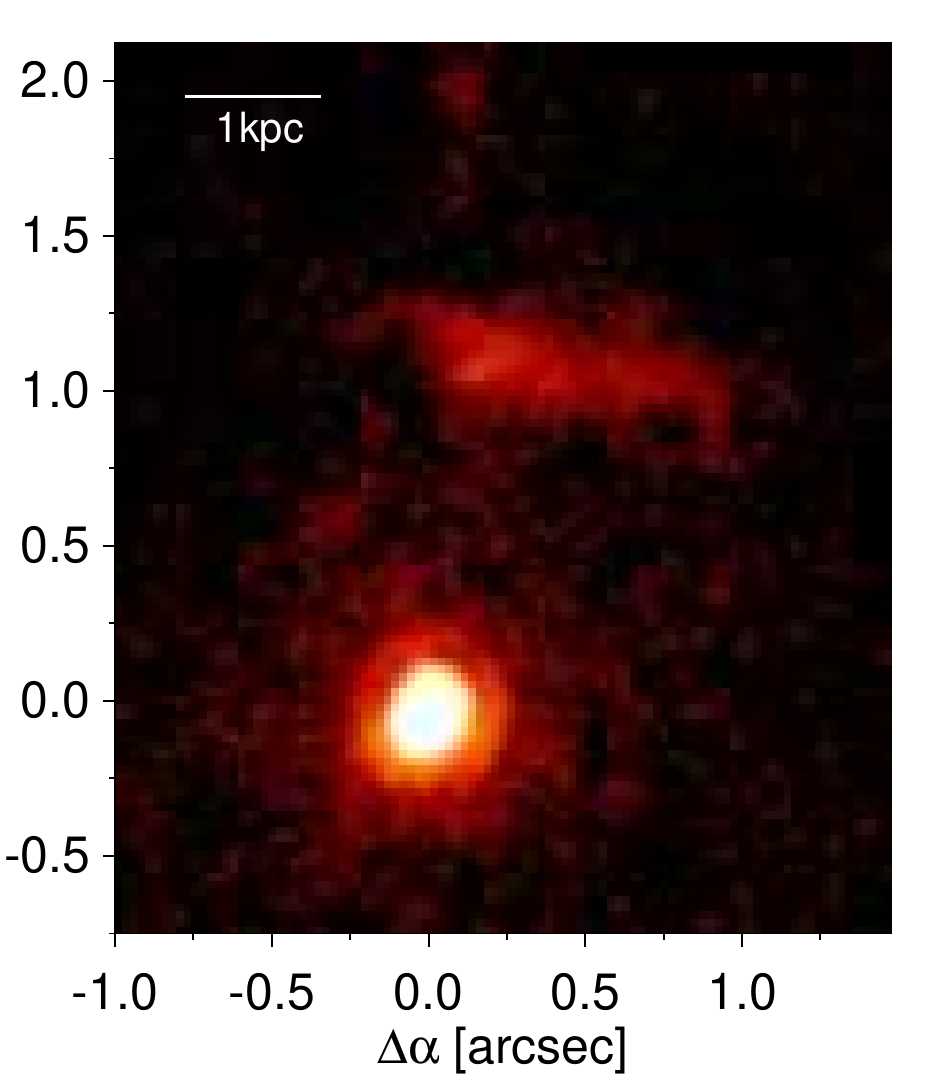}&
\hspace{-0.25cm}\includegraphics[width=0.24\textwidth]{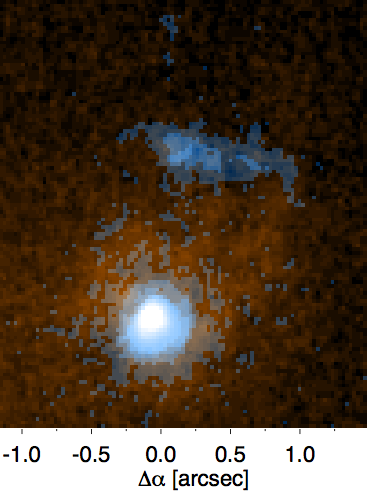}
\end{tabular}
\caption{Left to right: {\it HST}-ACS continuum, continuum subtracted [OIII],
  and overlay of continuum subtracted [OIII] (blue) on continuum (red) images of the 8 ULIRGs in our sample. North
  is up, East is left. The AGN nuclei are centred at 0,0. With the exception of F13428+5608 (Mrk273), the [OIII] emission is
  confined into a small region compared to the continuum emission from the
  galaxies. Therefore, the mid and right panels in the figure concentrate only
  in those regions with [OIII] emission, which are indicated with a dashed-line
  box in the corresponding continuum images. This Figure is best viewed in
  colour. Note that the logarithmic stretch used to display the images tends to emphasise the
extended structures relative to the compact nuclei. IRAS F14394+5332 is a
  merger between two galaxies that are separated 28 arcsec ($\sim$
  56\,kpc). Therefore, the two galaxies cannot be accommodated in the figure
  simultaneously. Since no [OIII] emission is detected from the western source,
  only the eastern source is included in this figure. The whole system is shown
  in Figure \ref{14394all} in Section 3.2.}
\label{HST_frames}
\end{figure*}

New {\it HST} ACS images for 7 ULIRGs in our sample were taken during Cycle 20 (GO:12934,
PI:C.N. Tadhunter). The Wide Field Channel (WFC, 0.050 arcsec pixel$^{-1}$) was
used in combination with narrow ($\sim$90$\AA$) and medium-band ($\sim$375$\AA$) filters. The narrow-band images
were positioned slightly blueward ($\sim$20$\AA$) of the narrow [OIII]$\lambda$5007 (hereafter \oxythr) emission
line in order to ensure that the blue wing of the \oxythr (representing the outflow), was in a wavelength region
of good sensitivity and also to include the [OIII]$\lambda$4959 emission. On the other hand, the medium-band filters were centered on the nearby line-free continuum
towards redder wavelengths. The sample was completed with IRAS\,F13451+1232 (PKS1345+12), which was observed during Cycle 13 using the
High Resolution Channel (HRC, 0.025 arcsec pixel$^{-1}$; \citealt{batcheldor07}). In this case, two medium-band
filters were used, centered on the \oxythr\  emission and the nearby continuum
towards bluer wavelengths. 

The data were reduced using the standard data reduction pipeline procedures. These
employ two packages: the {\sc CALACS} package, which includes dark
subtraction, bias subtraction and flat-field correction and produces calibrated
images, and the {\sc MULTIDRIZZLE} package, which corrects for distortion and
performs cosmic ray rejection. The remaining cosmic rays were removed manually
using the routines {\sc IMEDIT} in {\sc IRAF} and/or {\sc CLEAN} within the {\sc
  STARLINK} package {\sc FIGARO}.

To convert into physical units we used the PHOTFLAM header keyword. PHOTFLAM is
the sensitivity conversion factor and is defined as the mean flux density
F$_{\lambda}$ (in units of erg cm$^{-2}$ \AA$^{-1}$ counts$^{-1}$) that produces
1 count per second for a given {\it HST} observing mode. Since drizzled ACS
images are in units of counts s$^{-1}$, these were simply multiplied by the
PHOTFLAM value to obtain the flux in units of erg cm$^{-2}$ sec$^{-1}$
\AA$^{-1}$. The error associated with the flux calibration is $\sim$5\% for 
both the continuum and
the [OIII] images.

Once the images had been calibrated in flux, we subtracted the continuum from the
[OIII]~images to end up with  images that trace ``pure''
[OIII]~emission. Prior to that task, we used a series of {\sc IRAF} routines
to align the images. First, we used the
task {\sc SREGISTER} which registers an image to a reference image using
celestial coordinate information in the headers. Since three or more
foreground stars are always present in the ACS/WFC field of view, it
was then possible to use the tasks {\sc GEOMAP} and {\sc GEOTRAN} to geometrically
align the images and refine the final results. The accuracy of the alignment was
measured using the routine {\sc CENTER} that calculates the centroid of the
stars in the aligned images. We found that the images are aligned with an
accuracy better than 0.2 pixels. The continuum, continuum subtracted [OIII], and
[OIII]/continuum overlay images are shown in Figure \ref{HST_frames}.

In order to calibrate the [OIII]  images in terms of wavelength-integrated
[OIII]~emission-line flux (units: erg cm$^{-2}$ s$^{-1}$), an additional step was 
required. This involved multiplying the pixel values of the continuum-subtracted
[OIII] images by the effective bandwidths of
the ramp filters, as tabulated in Table 1. Note that the bandwidths of the filters
contain both the [OIII]$\lambda$5007 and the [OIII]$\lambda$4959 lines, so the derived
fluxes correspond to the sums of the fluxes of the two lines. The total [OIII]$\lambda$5007 emission line fluxes
measured using an aperture of 5\,kpc metrical diameter from the continuum-subtracted images
are shown in the penultimate column of Table 1. The latter have been corrected for the 25\% contribution
of the [OIII]$\lambda$4959 line within the filter bandpasses (assumed doublet ratio:
[OIII]5007/4959$=$3.0), and also for Galactic extinction using the V-band extinction estimates of 
\citet{schlafly11} available from the NASA/IPAC Extragalactic Database (NED). 

The final column of Table 1 shows a comparison between the HST-derived [OIII]$\lambda$5007 fluxes, and 
those measured by RZ13 from spectra extracted from WHT/ISIS long-slit data using extraction 
apertures with the same 5\,kpc length in the along-the-slit direction\footnote{Note that in the case of F13443+0832 the comparison is
made with the flux derived from the VLT/Xshooter data presented by Rose et al. (2017).}. Given that the shapes of the extraction apertures
are different (circular for the {\it HST} images, rectangular for the WHT/ISIS spectra), and that the ground-based data suffer from
seeing-related slit losses, it is not surprising that the fluxes derived from the ground-based spectra are generally lower than those
derived from the HST images. The discrepancy is largest in the case of the closest ULIRG in our sample (F13428+5608), as expected given that much of the emission in the 5\,kpc circular
aperture used to measure the HST flux this object falls outside the rectangular extraction aperture used by RZ13; the seeing measured 
at the time of the RZ13 observations of F13428+5608 was also worse that for most of the other objects
in the sample. However, in most cases the fluxes agree within 
40\%.

\addtocounter{figure}{-1}
\begin{figure*}
\centering
\begin{tabular}{ccc}
\hspace{-0.5cm}\includegraphics[width=0.34\textwidth]{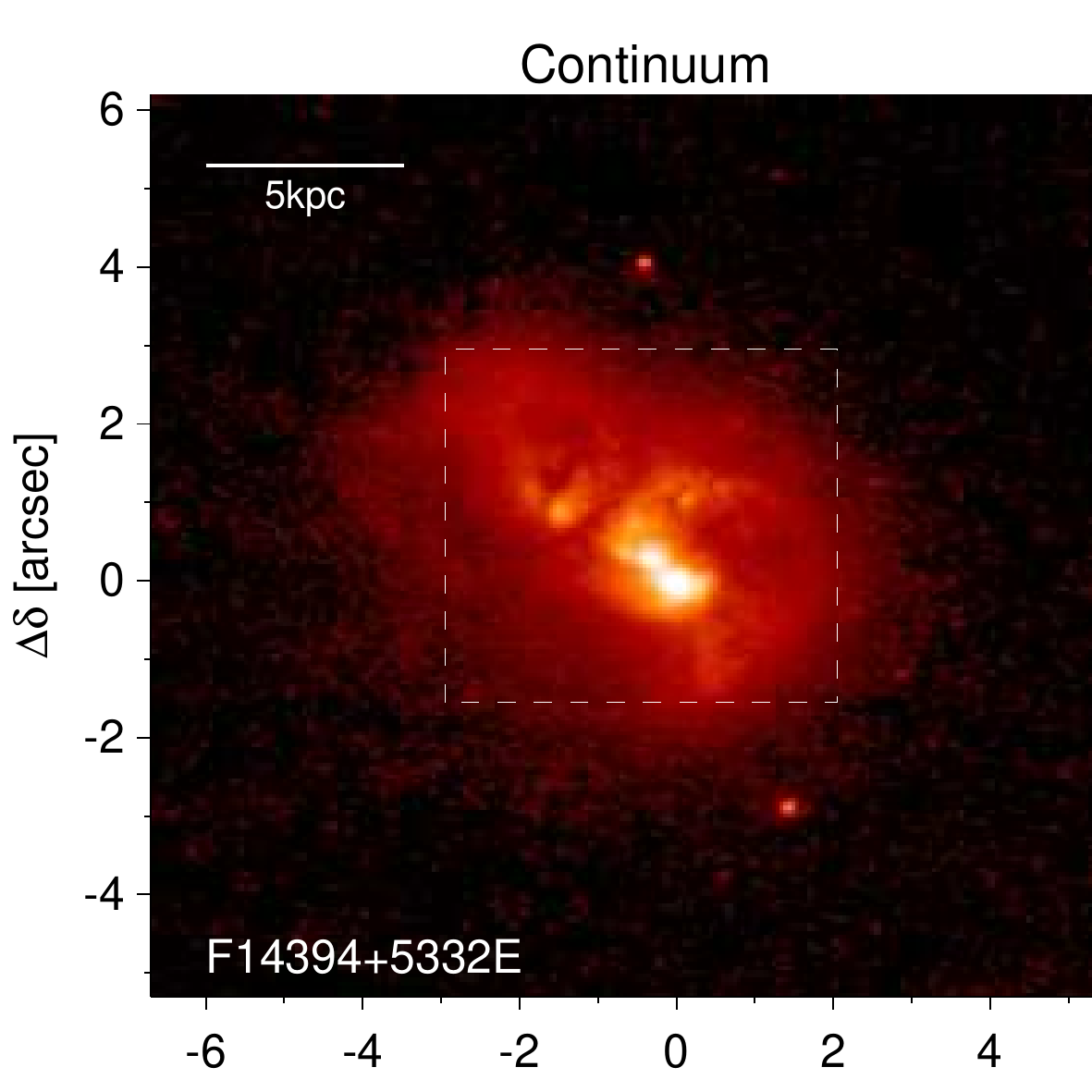}&
\hspace{-0.5cm}\includegraphics[width=0.342\textwidth]{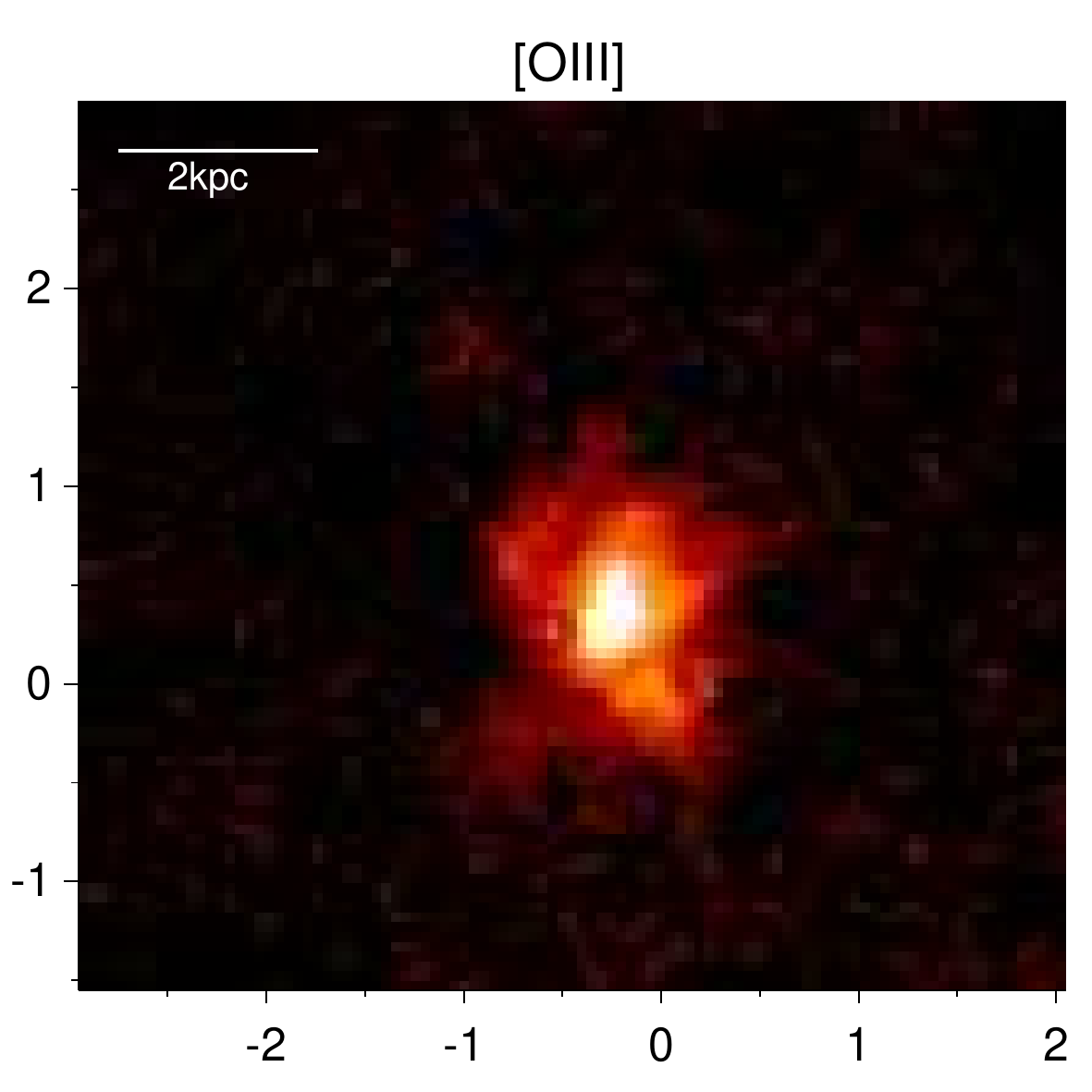}&
\hspace{-0.5cm}\includegraphics[width=0.315\textwidth]{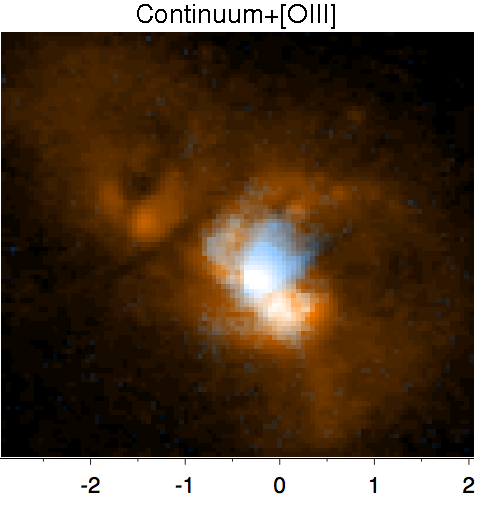}\\
\hspace{-0.5cm}\includegraphics[width=0.35\textwidth]{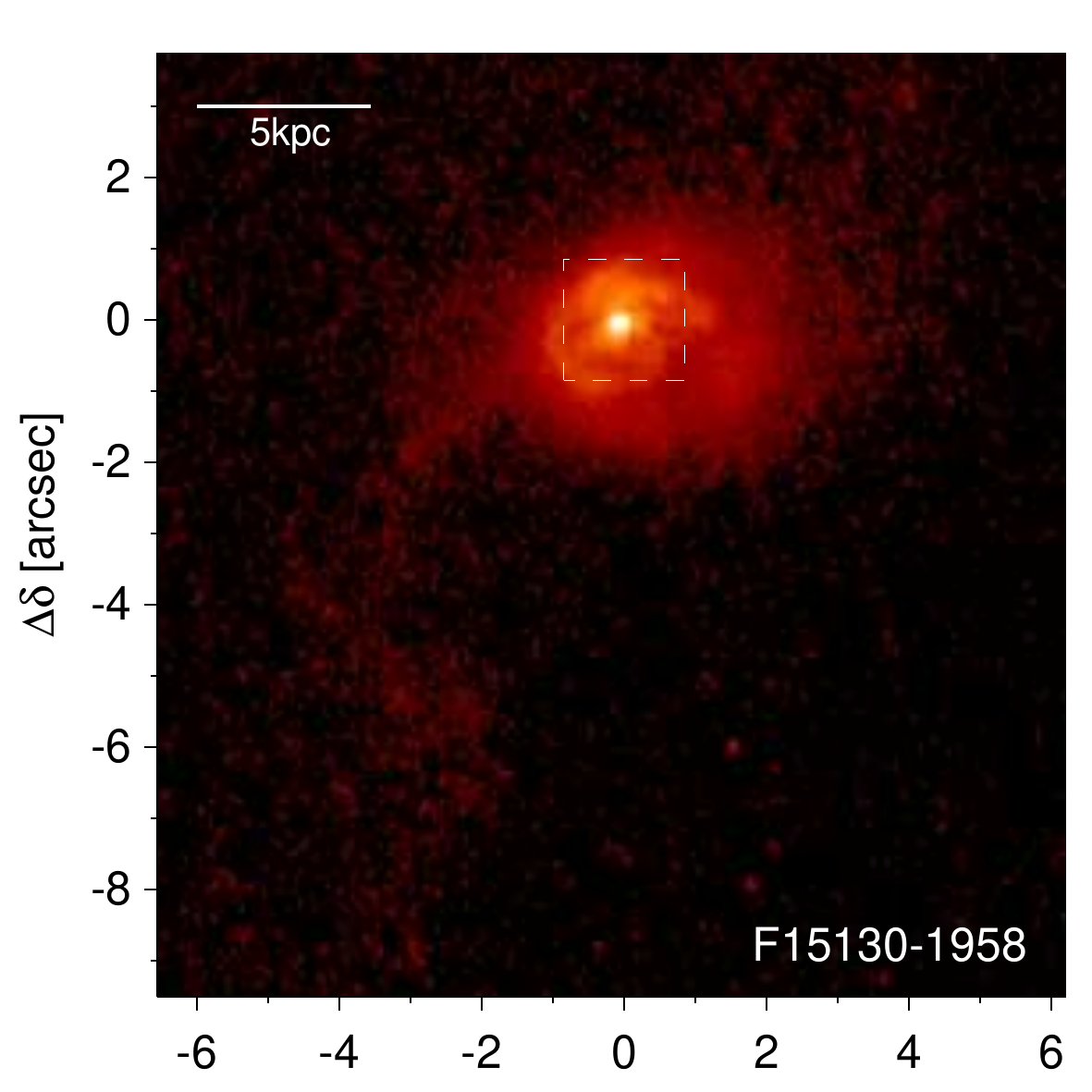}&
\hspace{-0.4cm}\includegraphics[width=0.345\textwidth]{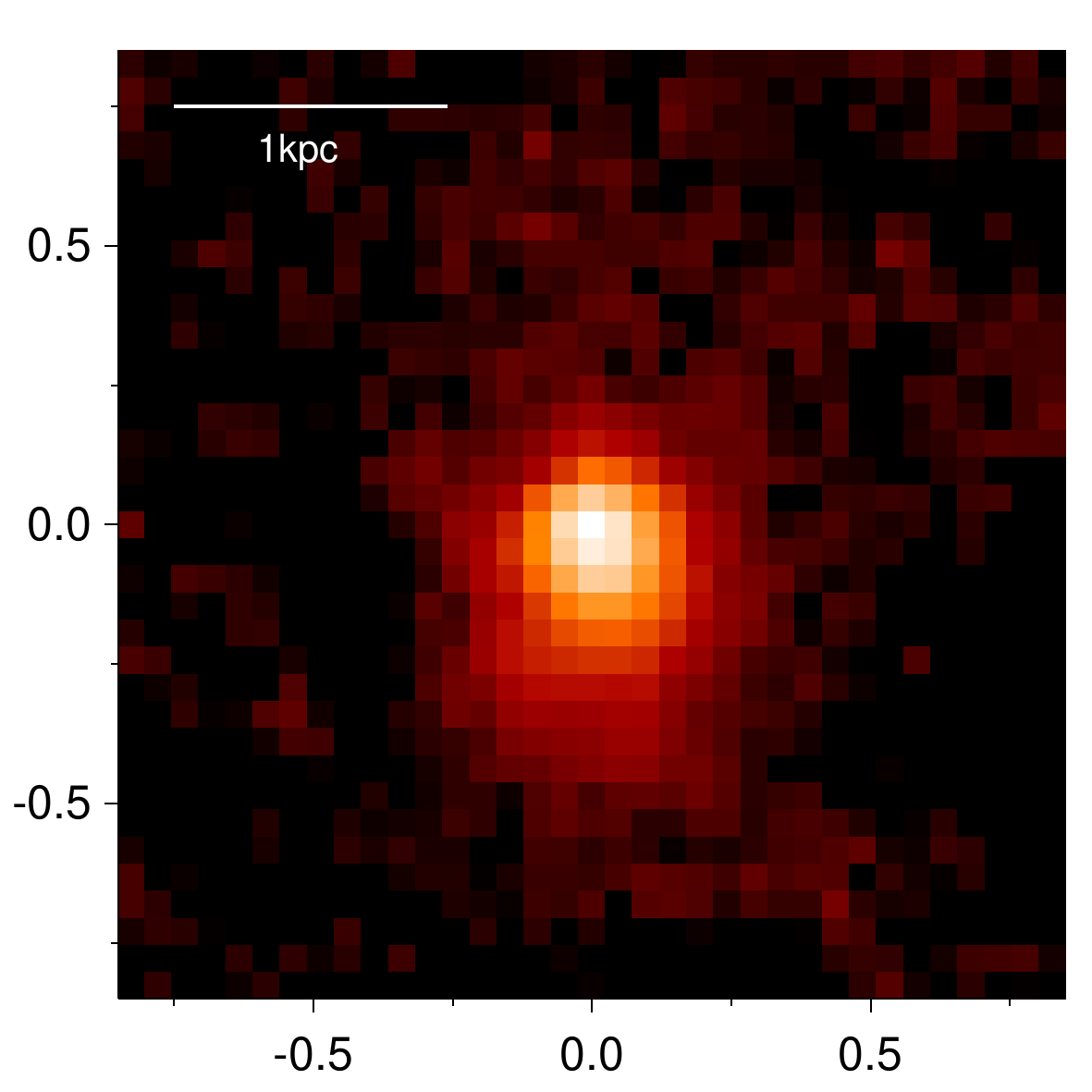}&
\hspace{-0.4cm}\includegraphics[width=0.30\textwidth]{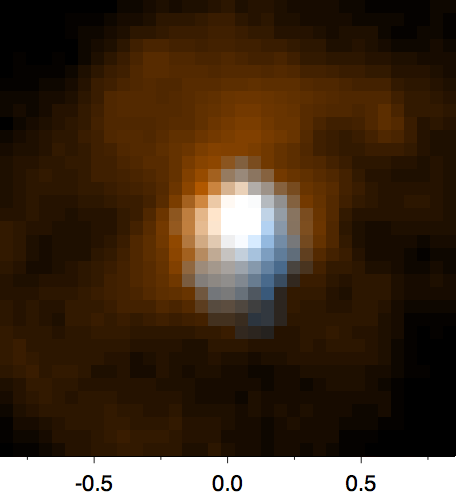}\\
\hspace{-0.3cm}\includegraphics[width=0.345\textwidth]{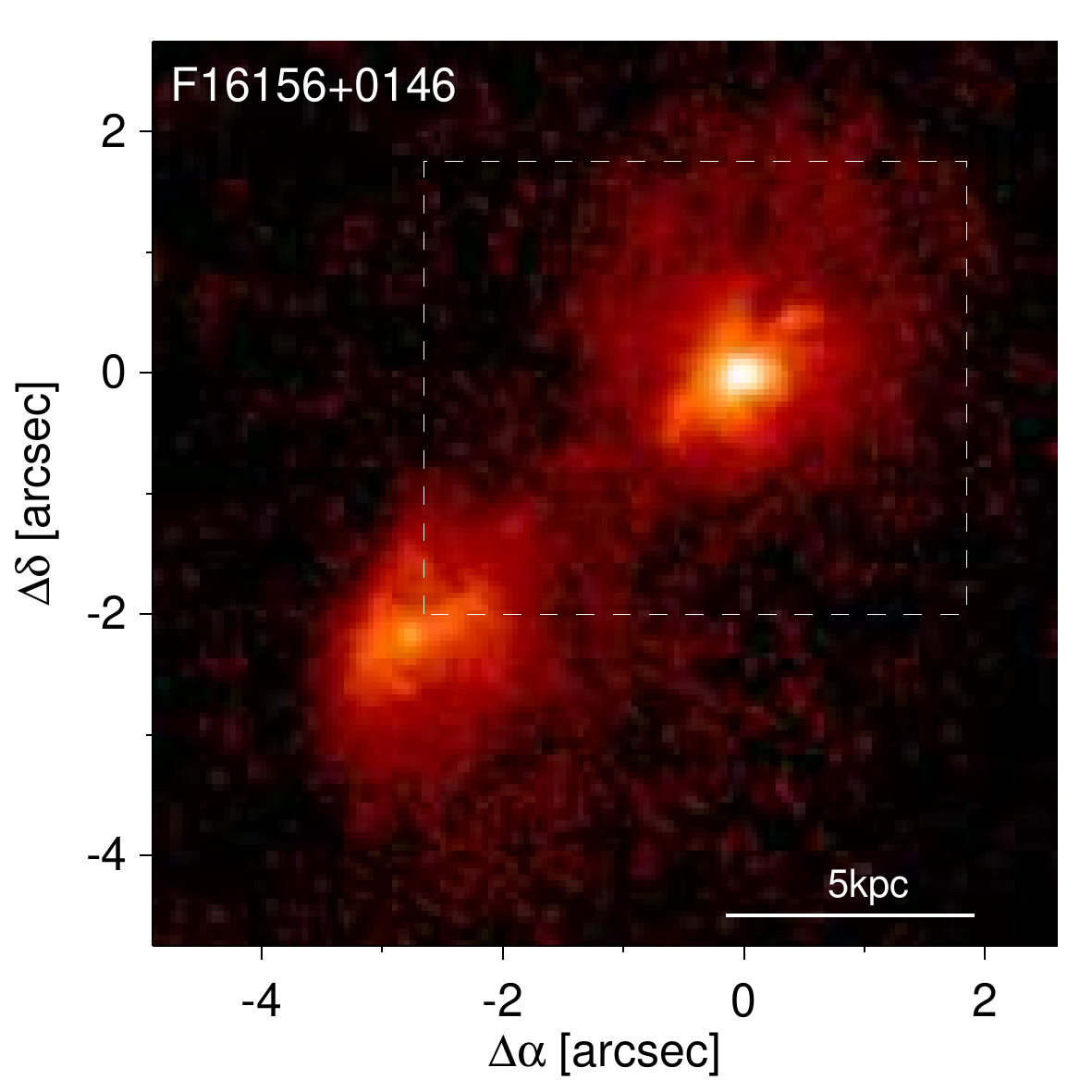}&
\hspace{-0.7cm}\includegraphics[width=0.35\textwidth]{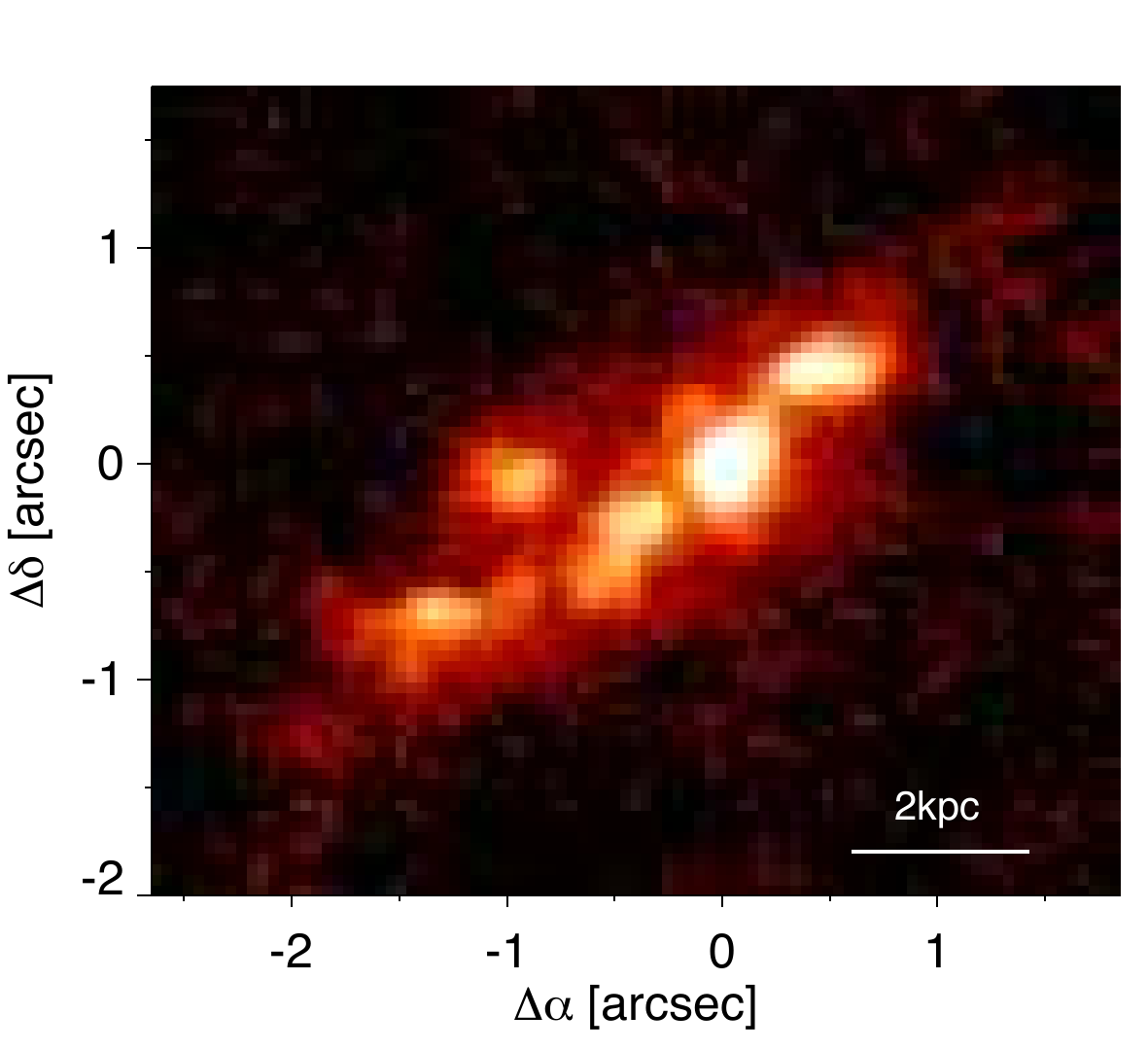}&
\hspace{-0.5cm}\includegraphics[width=0.312\textwidth]{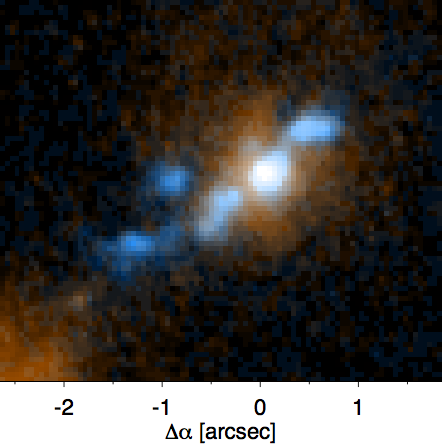}
\end{tabular}
\caption{Continued}
\end{figure*}

\section{Results}

It is clear from the images presented in Figure \ref{HST_frames} that the ULIRGs in our sample show considerable diversity
in their [OIII] emission-line morphologies, ranging from the spectacular arc structures observed on
scales of 10s of kpc in F13428+5608 (see also \citealt{rz14} and \citealt{spence16}), through
the complex near-nuclear structures that are resolved on scales of a few kpc in F13443+0802, F14394+5332, F16156+0146 and F17044+6720, to 
compact structures that dominate the nuclear emission in F1345+1232, F15130+1958, and F17179+544. Note that,
although many of the objects show spectacular extended emission-line structures, these extended structures are not necessarily associated with AGN-driven outflows (see discussion in section
3.2 below).  Before
considering each object in detail, in the next section we describe
the measurement of the radii of the warm outflow regions in the target objects.

\subsection{Measuring the radii of the outflow regions}\label{sect:main}

Although the morphologies of the [OIII]-emitting regions are diverse in our ULIRG sample, five of the sources contain compact [OIII]-emitting regions centred on the continuum nuclei; in three of these five cases, these compact nuclear regions contribute the bulk of the [OIII] emission on the scales of the 5\,kpc diameter apertures used for the ground-based spectroscopic observations of RZ13 (see below). Therefore, it is important to determine whether these compact nuclear emission-line regions are spatially resolved and, if so, to determine their radii.


We have estimated the radii of the compact, near-nuclear [OIII] structures (R$_{[OIII]}$) by fitting 2D Gaussians to the central cores seen in the [OIII] continuum-subtracted HST/ACS images. The FWHM of the 2D Gaussians were measured using the {\sc IRAF} routine {\sc imexamine}. We measured the FWHM for both the compact nuclear [OIII] components and stars in the field. Note that for the stars the FWHM were measured in the continuum images, which have a higher S/N for faint stellar images. 

Given that the compact nuclei and stars are all detected at high signal-to-noise, much of the uncertainty in the FWHM measurements is likely to be related to the fact that the ACS detector pixels under-sample the point spread function (psf) at these wavelengths, particularly in the case of the WFC observations used for most of the objects. Therefore, we estimated the uncertainly in the FWHM measurements by resampling the images multiple times and re-fitting the compact [OIII] emission and stars in each resampled image. The resampling was affected by shifting the images by non-integer pixel amounts in 8 different directions. A total of 24 re-samplings were carried out for each object frame, and the uncertainty on the FWHM for the compact [OIII] sources and the stars was taken as the standard error in the mean of the fitted FWHM measurements for all the resampled images. The results are shown in Table \ref{table:2dgauss}, where the mean FWHM measurements for the compact nuclear [OIII] sources are listed along with those for stars measured in the continuum images. 

\addtocounter{figure}{-1}
\begin{figure*}
\centering
\begin{tabular}{ccc}
\hspace{-1.5cm}\includegraphics[width=0.34\textwidth]{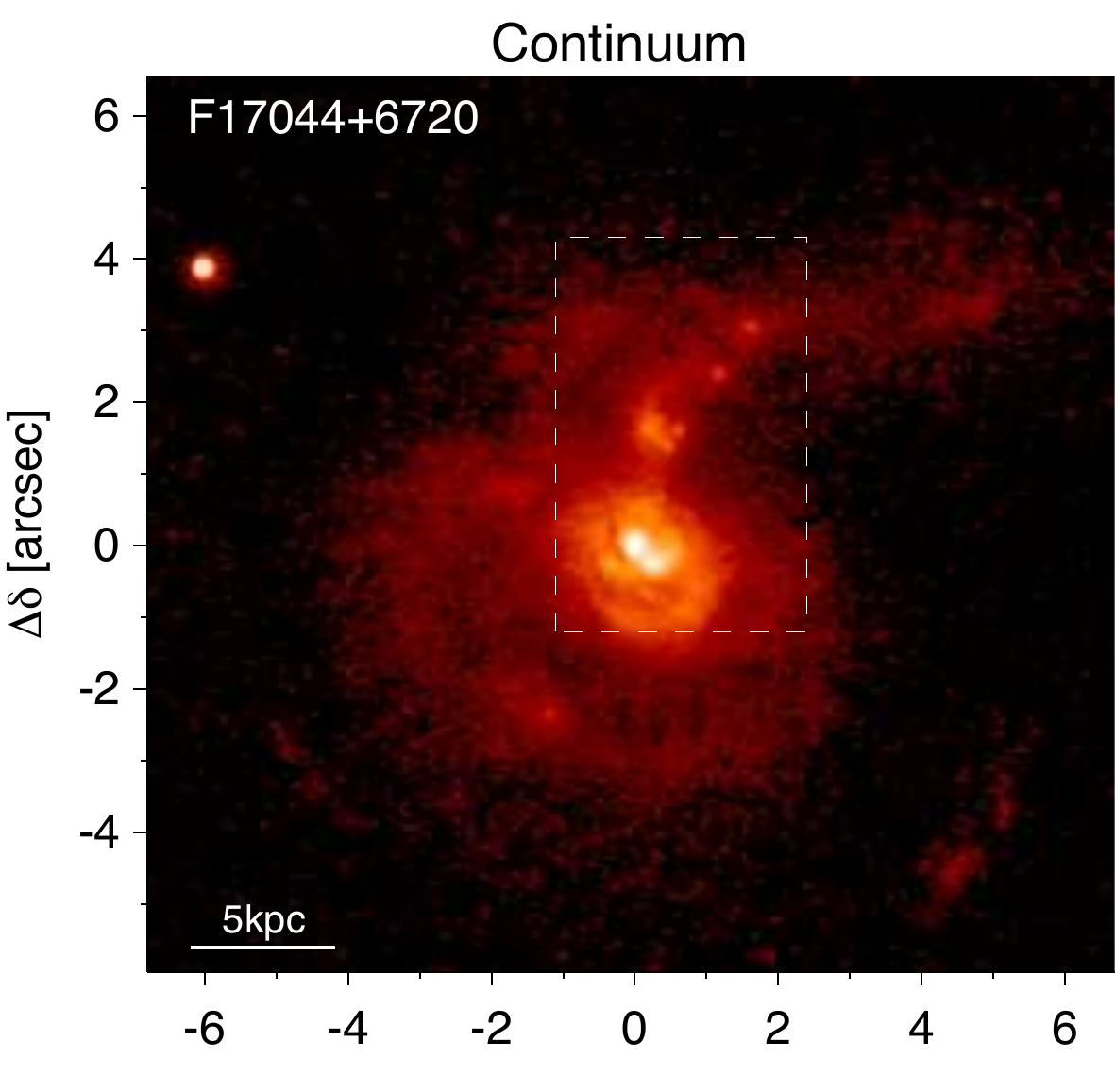}&
\hspace{-0.cm}\includegraphics[width=0.2\textwidth]{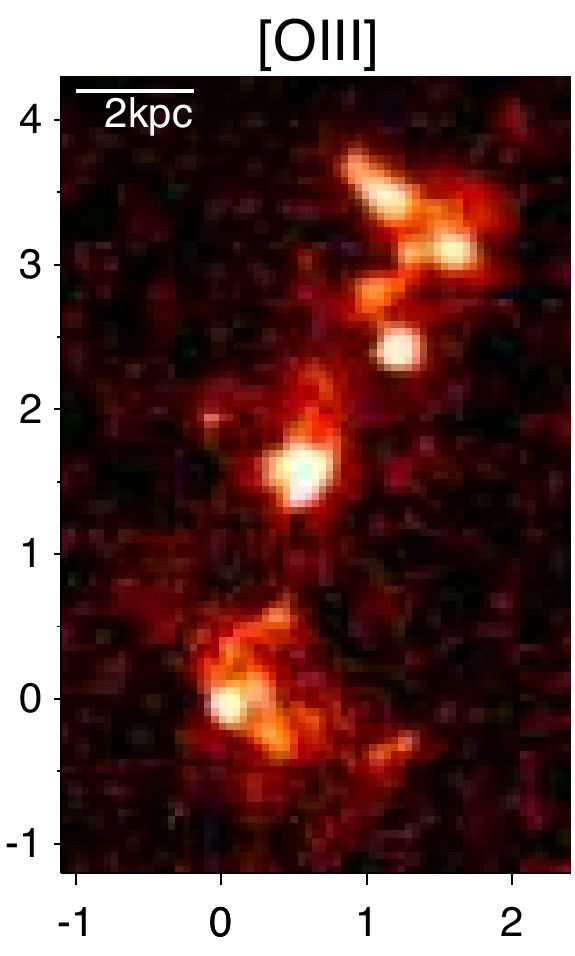}&
\hspace{-0.2cm}\includegraphics[width=0.1775\textwidth]{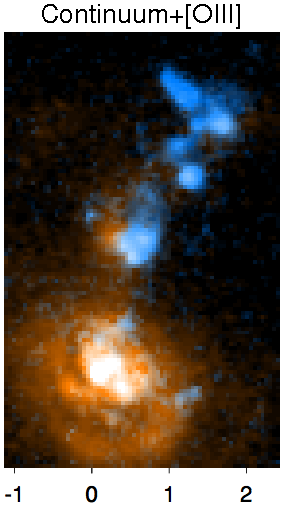}\\
\hspace{-1.4cm}\includegraphics[width=0.36\textwidth]{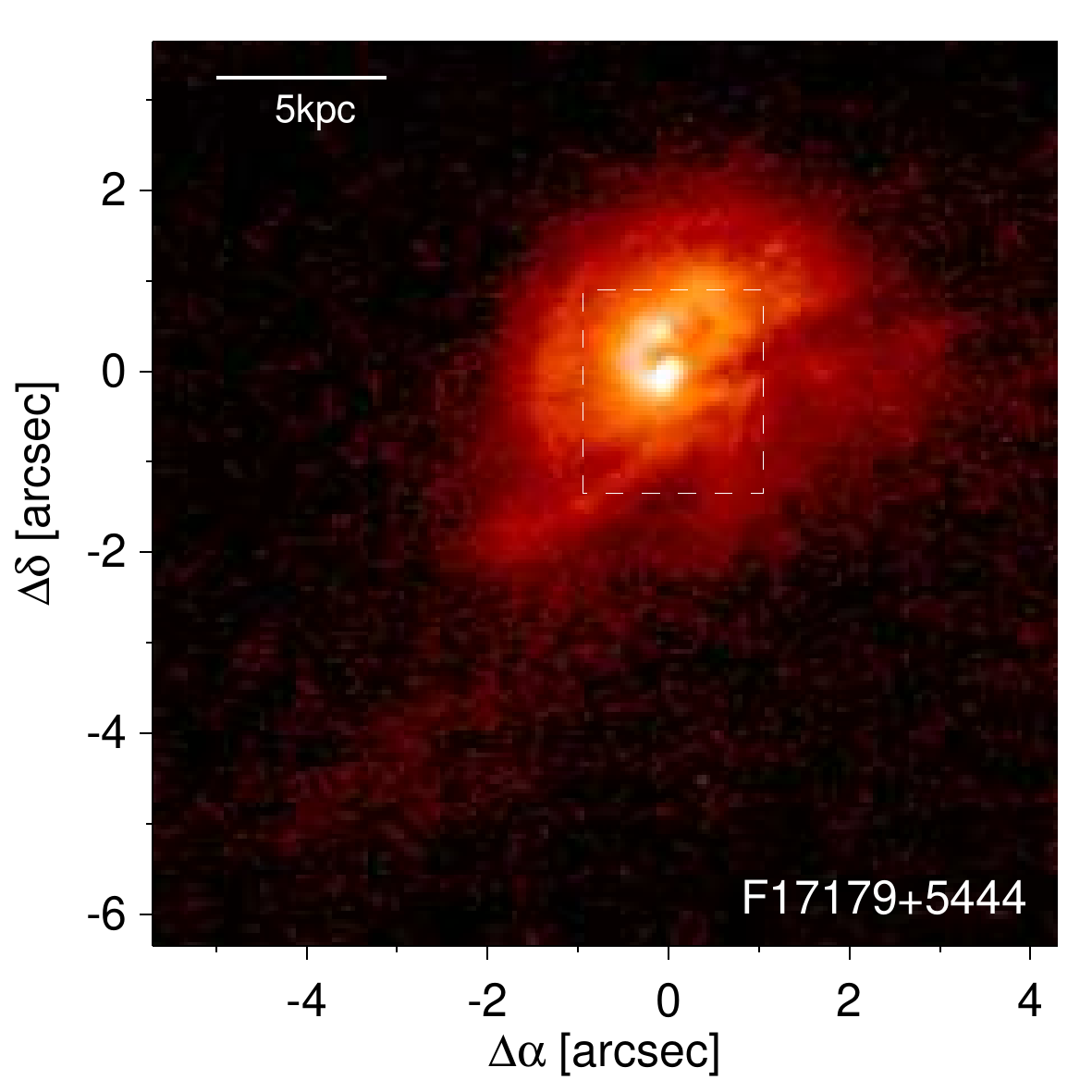}&
\hspace{-0.3cm}\includegraphics[width=0.305\textwidth]{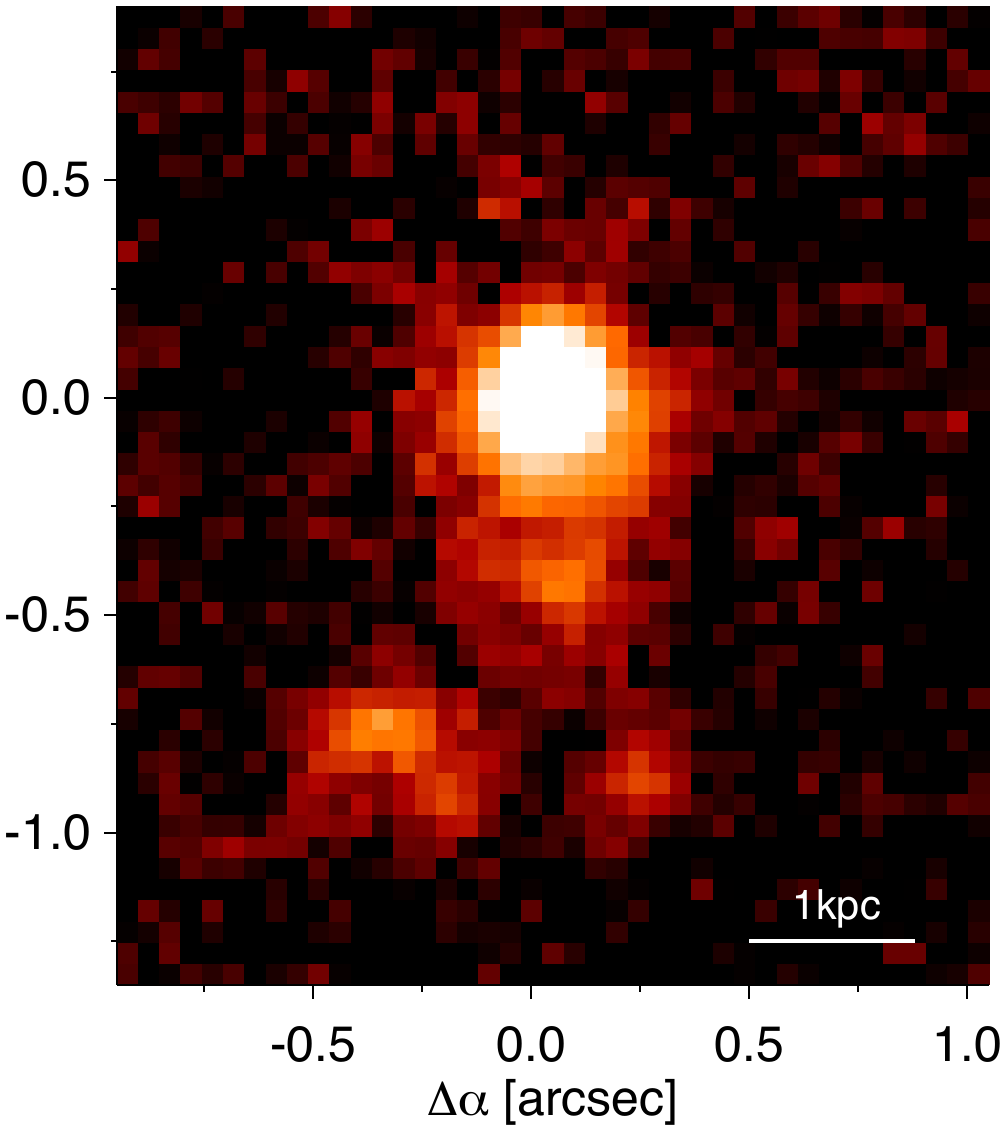}&
\hspace{-0.3cm}\includegraphics[width=0.27\textwidth]{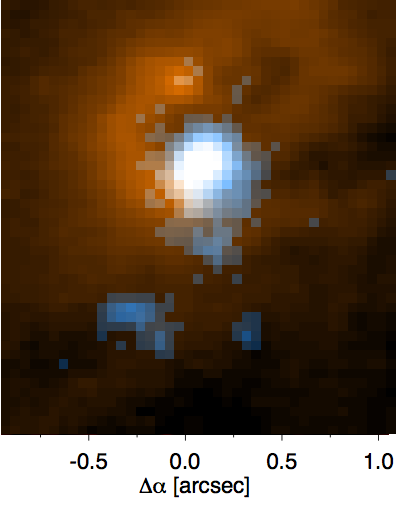}
\end{tabular}
\caption{Continued}
\end{figure*}

Comparing the [OIII] FWHM with those measured for the stars, it is clear that the compact [OIII] emission line regions are spatially resolved in all five of the objects listed in Table \ref{table:2dgauss}, in the sense that the [OIII] FWHM are significantly larger than those of the stars (the differences between the compact [OIII] and stellar measurement are $>5\sigma$ in all cases, where $\sigma$ is the estimated uncertainty in the difference). The penultimate column of Table \ref{table:2dgauss} gives the instrumentally corrected radii of the compact nuclear [OIII] regions. These were estimated under the assumption that the 2D light distributions of both the compact [OIII] components and the stars are Gaussians, with the instrumentally-corrected [OIII] FWHM  obtained by subtracting the stellar FWHM from the measured [OIII] FWHM in quadrature. The radii of the compact [OIII] regions were then taken as equivalent to half the instrumentally-corrected FWHM
(i.e. the HWHM). 

It is important to note that the uncertainties
listed for
the [OIII] outflow radii in Table \ref{table:2dgauss}  -- based on the results of the multiple re-samplings of the images described above -- are likely to underestimate
the true uncertainties. This is because we have assumed that spatial profiles of both the star and the [OIII] flux distributions are perfectly represented by 2D Gaussians, whereas in reality this is only an approximation.


In the cases for which there was no compact nuclear source (F13443+0802, F14394+5332, F17044+6720), or for which the extended [OIII] emission makes a substantial contribution within the RZ13 spectroscopic 
aperture (F13428+5608, F16156+0146), we have 
estimated a flux-weighted mean radius for the emission inside a 5\,kpc diameter aperture centred on the nucleus. The results are shown in Table \ref{table:fmean}. Note that in the case of F14394+5332 there is some ambiguity about the true position of the nucleus, so we give two values that correspond to two different assumptions about the nuclear position. 

Finally, we also give estimates of the maximum radii of the outflows in the final columns of Tables \ref{table:2dgauss} \&
\ref{table:fmean}. In cases where we have spectroscopic information about the kinematics of any extended emission beyond the 5\,kpc nuclear aperture of RZ13, we use this information to determine the maximum outflow radius. For all other objects, the maximum radius is set to the maximum extent of the [OIII] emission visible in the {\it HST} images. In such cases, the maximum radius may over-estimate the true outflow radius, because the gas  emitting the [OIII] emission outside the 5\,kpc spectroscopic aperture may not be taking part in the outflow, even if it is photoionized by the AGN.


\begin{table*}
\centering
\begin{tabular}{llllll}
\hline\hline
IRAS FSC &FWHM$^{meas}_{[OIII]}$ &FWHM$_{stars}$  &HWHM$^{corr}_{OIII}$ &R$_{[OIII]}$ &R$^{max}$ \\
&(pixels) &(pixels) &(arcsec) &(pc) &(kpc) \\
\hline
F13428+5608  &3.17$\pm$0.01  &2.44$\pm$0.05 &0.051$\pm$0.002 &37$\pm$2 &5  \\
F13451+1232  &3.75$\pm$0.01	&2.70$\pm$0.01  &0.0351$\pm$0.00025 &69$\pm$1 &3.3 \\
F15130+1958  &2.71$\pm$0.04	&2.34$\pm$0.005	&0.034$^{+0.004}_{-0.005}$ &66$^{+8}_{-10}$ &0.77 \\
F16156+0146  &2.88$\pm$0.04  &2.43$\pm$0.03 &0.038$\pm$0.03 &87$\pm$7 &4.3 \\
F17179+5444  &2.56$\pm$0.04	&1.81$\pm$0.03	&0.046$\pm$0.003 &112$\pm$7 &2.6 \\
\end{tabular}
\caption{Measurements of the sizes of the compact [OIII]-emitting regions in the
nuclei of the target objects.  Col (1): object designation in the IRAS Faint Source Catalogue
  Database. Col (2): FWHM of the 2D Gaussians of the compact [OIII] cores in pixels. Col (3): FWHM of the 2D Gaussians of the stars in the field in pixels. Col (4): HWHM of the compact cores corrected for the instrumental width in arcseconds. Col (5): the estimated
radius of the outflow region derived from the HWHM in pc. 
Col (6): estimate of the maximum possible outflow radius that
is consistent with existing {\it HST} imaging and ground-based spectroscopy.  Note that the pixel scale for the PKS1345+12
(0.025 arcsec pixel$^{-1}$) is smaller than that for the other objects (0.05 arcsec pixel$^{-1}$) because it was taken with the ACS/HRC rather than the ACS/WFC.}
\label{table:2dgauss}
\end{table*}




Given the wide range of emission line morphologies, we now discuss the spatial extent of the warm outflow region for each object in turn. For most cases, the main challenge we face is that the ground-based spectra that were used to detect the outflows by examining the emission line profiles have a much lower spatial resolution than the {\it HST} imaging observations. However, the fact that the outflow components dominate the nuclear [OIII] emission line profiles in many cases allows us to draw strong conclusions about the radii of the warm outflows.

\subsection{Descriptions of individual objects}

In this section we describe the continuum and [OIII] morphologies 
for the 8 ULIRGs in our sample shown in Figure \ref{HST_frames}, and use the 
imaging data, combined with the existing ground-based spectroscopy results, to estimate the radial extents of the outflow regions.

\subsubsection{F13428+5608 (Mrk273)} 

At a redshift of only $z=0.037$, the Mrk273 system is one of the closest, and best resolved, of all ULIRGs. Recently,  imaging and spectroscopic observations of
Mrk273 have been  presented by \citet{rz14} and \citet{spence16}. The deep {\it HST} [OIII] images presented in \citet{rz14} and Figure 1 reveal  a spectacular structure of ionised gas in the form of arc-like filaments and clumps that extend 25 arcseconds ($\sim$19\,kpc) to the east of the nuclear
region, in a direction that is close to perpendicular to prominent tidal tail to the south. The full radial extent of the warm ionized gas, as revealed by deep H$\alpha$ imaging observations with the Gran Telescopio Canarias (GTC), is $\sim$45\,kpc \citep{spence16}. Based on the overall morphology of the extended continuum structures, as well as the likely presence of three nuclei in the central regions, \citet{rz14} have suggested that this is object represents a triple merger.

Despite its large radial extent, the existing long-slit spectra of the very extended ionized nebula reveal that
it is kinematically quiescent, with relatively small line widths ($FWHM \le 350$~km s$^{-1}$) and velocity shifts ($\Delta V < 250$~km s$^{-1}$) at radial distances greater than 6\,kpc from the nucleus \citep{rupke13,rz14,spence16}. In contrast,  various long-slit and integral field observations of the ionized gas on smaller scales find evidence for a high degree of kinematic disturbance with linewdiths $FWHM > 1,000$~km s$^{-1}$ in some near-nuclear regions \citep{colina98,rupke13,rz14}. Indeed, this is the only ULIRG object in the current sample in which the  near-nuclear outflow is well resolved in ground-based spectroscopy observations.  The maximum radial extent measured for the outflow in the ground-based spectra depends on how we define an outflow: if we use the criterion of this paper (i.e. an outflow is any component with $FWHM > 500$~km s$^{-1}$), the maximum (spectroscopic) radial extent of the outflow is 
5\,kpc; however, even if we use the more conservative criterion that an outflow is associated with larger line widths $FWHM > 1,000$~km s$^{-1}$, the maximum radial extent is only  marginally smaller ($\sim$4\,kpc). Therefore we take 5\,kpc as the maximum extent of the near-nuclear outflow in this case. Aside from questions about how we define an outflow, it is important to add the caveat that the outflows are not necessarily AGN-driven out to the full radial extent of 5\,kpc: given the general weakness of the detected AGN activity in this case (RZ13), starburst-driven outflows may also make a significant contribution.

Although the outflow in Mrk273 is clearly resolved in ground-based spectra, for completeness and comparison with the other objects in our sample, in Table \ref{table:2dgauss} we present estimates of radial extent of the compact [OIII] emission associated with the SE nucleus \citep[see][]{rz14}. This compact nucleus contributes only $\sim$7\% of the
total [OIII] flux on the scale of the spectroscopic aperture used by RZ13 to study the near-nuclear outflow. For comparison, the broader of the two components detected in this nuclear aperture ($FWHM =1368 \pm 66$~km s$^{-1}$) contributes 40\% of the total [OIII] flux (RZ13). Therefore, the compact emission-line region associated with the SE nucleus is unlikely to dominate the outflow in this case. 

\begin{table}
\centering
\begin{tabular}{lcc}
\hline\hline
IRAS FSC &Flux-weighted &R$^{max}$ \\
         &radius (pc)   &(kpc) \\
\hline
F13428+5608  &$1801\pm3$   &5 \\
F13443+0802  &$782\pm8$  &1.9 \\
F14394+5332P  &$652\pm6$  &1.9	\\
F14394+5332D  &$840\pm8$     &2.2    \\
F16156+0146  &$849\pm4$  &4.3  \\
F17044+6720  &$1184\pm6$  &3.2   \\
\end{tabular}
\caption{Measurements of the radial extents of the nuclear [OIII]-emitting regions for objects
in which the nuclear regions are clearly resolved within the apertures used for the
spectroscopy.  Col (1): name. Col (2): Flux weighted mean radius of the [OIII]-emitting region, as measured within
a 5\,kpc diameter aperture centred on the assumed
position of the true nucleus containing the supermassive black hole. Col (3): the maximum outflow radius that
is consistent with existing {\it HST} imaging and ground-based spectroscopy observations. Note that in
the case of F14394+5332 two sets of values are given because there is some ambiguity about
the true position of the nucleus: F14394+5332P represents the case in which the AGN nucleus is centred on the peak
of the [OIII] emission in the more northerly continuum condensation in the E component, whereas F14394+5332D
represents the case in which the AGN nucleus is centred in the dust lane between the two continuum condensations.
For F13428-5608 the flux weighted mean radius was measured relative
to the SE nucleus associated with the compact [OIII] structure (see RZ13).}
\label{table:fmean}
\end{table}

\subsubsection{IRAS F13443+0802} 

IRAS F13443+0802 is a spectacular triple system comprising a
close pair with a nuclear separation of $\sim$5 arcsec ($\sim$12\,kpc) in the N-S direction, along with a
third galaxy located $\sim$15 arcsec ($\sim$35\,kpc) to the SW of that pair. We
will refer to the individual galaxies as the NE, SE, and SW components
respectively (following the notation of Veilleux et al. 2002). 
The {\it HST} continuum image  shows a highly
disturbed morphology in the case of the NE component. Several clumps and
condensations, along with dust features, are observed through the entire body of
the galaxy, but are more prominent in the nuclear region and towards the south
of this source. Overall, the NE component has the appearance of a distorted spiral galaxy
observed close to face-on. 

In contrast, the SE component presents an elongated continuum morphology, extended from E to W along PA
$\sim$ 81$^{\circ}$. In addition, a prominent dust lane  crosses the western
part of the galaxy with the same PA as the overall elongation of the continuum emission, 
which suggests that we are 
observing an edge-on or highly inclined disk. A bridge of faint, diffuse
continuum emission is visible between the SE and the SW component, with the latter
showing a regular spiral morphology in continuum emission.


We note that there is some ambiguity surrounding the identity of the AGN nucleus in this system.
\citet{veilleux99} found no evidence for significant [OIII] emission from the NE nucleus —- consistent with our {\it HST} imaging
results -- but found that the SW component
has emission line ratios consistent with a Sy2 classification; the SE component is not discussed in the latter paper. However,
our {\it HST} [OIII] images show strong, extended [OIII] emission from the SE component. This ambiguity is resolved by our VLT/Xshooter observations \citep{rose17} which detect strong, high-ionisation emission lines (e.g. HeII$\lambda$4686, [NeV]$\lambda$3426 in the nucleus of the SE galaxy which are not present in
the spectrum of the nucleus of the SW galaxy, thus demonstrating that the dominant AGN component is associated with the SE galaxy.


However, our most interesting finding for IRAS F13443+0802 is related to the morphology of the SE
galaxy. The [OIII] emission from this galaxy has a peculiar ``Y''
morphology, centred
on the region of brightest continuum emission  along the disk, and bisected by the prominent
dust lane. The overall
morphologies of the continuum and [OIII] emission of the SE component suggest
that the Y structure corresponds to the two sides of an emerging
outflow in an edge-on or highly inclined system. 

The extraction aperture  used for our VLT/Xshooter spectrum of F13443+0802SE \citep{rose17} encompasses all the [OIII] emission from the high-surface-brightness Y structure.  In order to adequately fit the [OIII] emission line, three Gaussian components are required: two  narrow ($FWHM = 350 \pm 6$ and $73\pm14$ km s$^{-1}$ ) components and one intermediate ($FWHM = 698 \pm 42$ km s$^{-1}$ ) component, shifted by $\Delta V = +60\pm48$,  $-57\pm45$ and $-50\pm46$ km s$^{-1}$ respectively relative to the rest frame defined by the stellar absorption features. The 
presence of the intermediate component -- which contributes 53\% of the total [OIII] flux -- provides evidence for kinematic disturbance in the emission line gas. Note that, in this case,
the relatively small velocity shifts and widths of the kinematic components relative to the host galaxy rest frame could be explained by the fact that the system is observed close to edge-on.

If we assume that the AGN nucleus of IRAS F13443+0802~SE lies behind the dust lane that crosses the Y structure, the maximum radial extent of the near-nuclear outflow measured from our [OIII] image is 0.8 arseconds (1.9\,kpc), and the flux-weighted mean radial extent of the outflowing gas in the Y structure is 0.34 arcsec (0.78\,kpc).




\subsubsection{IRAS F13451+1232 (PKS1345+12)} 

IRAS F13451+1232 (PKS1345+12) is a well studied object that shows evidence for outflows in warm, neutral and molecular
gas \citep{holt03,morganti05,holt11,dasyra12,morganti13}. Deep {\it HST} imaging and ground-based spectra for PKS1345+12 are discussed in \citet{holt03}, \citet{batcheldor07} and
\cite{holt11}. We note that the AGN in this object is radio-loud ($P_{1.4GHz} = 1.9\times10^{26}$ W Hz$^{-1}$ m$^{-2}$) with a compact, high-surface-brightness inner radio 
structure that includes a jet that has a maximum radial extent of $\sim$130pc to the S along PA180 \citep{morganti13}, as well as a more diffuse
extended radio structure with a total extent of $\sim$150\,kpc in a roughly N-S direction \citep{stanghellini05}. Given the similarity between the spatial extent of the  inner radio structures and those of the warm, neutral and molecular outflows in this source,  there is a strong possibility that the outflows are driven by the relativistic jets (see the discussion in \citealt{batcheldor07} and \citealt{morganti13}).

The {\it HST} continuum image shown in Figure 1 is dominated by a double nucleus of separation 2.1 arcsec (4.5\,kpc) which represents the cores of the two merging galaxies in the system. The western nucleus contains the powerful, radio-loud AGN, whereas the eastern nucleus is quiescent.

The deep {\it HST} [OIII] image of F13451+1232 has a better resolution and spatial sampling than those of the other objets in the sample, because it was taken with the high resolution camera (HRC) rather than the WFC of the ACS instrument.  Even with this better spatial resolution, the  nuclear [OIII] emission of F13451+1232 is barely resolved: although the emission is slightly elongated roughly in the direction of the radio jets, the first Airy ring is visible in the image. In addition to the compact core, an arc feature is detected $\sim$1.2 arsec ($\sim$2.5\,kpc) to the NW of the nucleus in the [OIII] image. The instrumentally-corrected HWHM of the compact core source is $0.069\pm0.001$\,kpc. 

The spectroscopic observations of Holt et al. (2003: 1.3$\times$2.1 arcsec aperture along PA160) encompass the entire compact [OIII] structure, but miss much of the emission of the prominent extended arc $\sim$1.2 arcseconds to the north of the nucleus. Within this spectroscopic aperture the [OIII] emission lines show complex profiles which can be modelled as a narrow component ($FWHM=340\pm23$ km~s$^{-1}$), an intermediate component ($FWHM=1255\pm12$ km~s$^{-1}$) and a broad component ($FWHM=1944\pm65$ km~s$^{-1}$), with the intermediate and broad components shifted by $-402\pm9$ and $-1980\pm36$~km~s$^{-1}$ respectively relative to the narrow component; this is one of the most kinematically disturbed objects in our sample.
The broad and intermediate components --- which we take to represent the outflow --- contribute 95\% of the total [OIII] flux in the spectroscopic aperture \citep{holt03}. 

Because of uncertainties about the seeing FWHM for the spectroscopic observations we
do not directly compare the absolute fluxes of the [OIII]$\lambda$5007 emission between the spectroscopic and imaging observations. Rather, we base our analysis of the spatial extent of the outflow on a comparison between the proportion of the total [OIII] emission that is emitted by the compact nuclear source, as derived from the imaging observations, and the proportion of total [OIII] flux emitted by the broad and intermediate outflow components derived from the spectroscopic observations.

First, we estimate the total [OIII] flux emitted by the compact nuclear source by performing photometry on our HST/ACS image: within a circular aperture of radius 0.314 arcsec (12.5 pixels) that is dominated by the compact source, and measure an [OIII] flux of $6.0\times10^{-14}$ erg s$^{-1}$ cm$^{-2}$. We then apply an aperture correction of factor 1.112 under the assumption that the compact source has a spatial profile similar to that of the  psf of the {\it HST ACS/HRC} at the wavelength of the filter\footnote{Given the compact nature of the high-surface-brightness structure and detection of the first Airy ring, this is a reasonable approximation. However, it might lead to a slight underestimation of
the aperture correction because the compact nuclear source appears slightly elongated.} \citep{sirianni05}. In this way, we estimate that the compact source has a total [OIII] flux of $7.4\times10^{-14}$ erg s$^{-1}$ cm$^{-2}$. In comparison, the total [OIII] flux within a circular aperture of radius  1.05 arcsec --- corresponding to the radial extent of the nuclear spectroscopic extraction aperture --- is $9.1\times10^{-14}$ erg s$^{-1}$ cm$^{-2}$. Therefore, the compact nucleus contributes 81\% of the total [OIII] flux in the spectroscopic aperture -- similar to the proportion of the total [OIII] emission contributed by the broad and intermediate outflow components in the same aperture.


To investigate the total extent of the near-nuclear outflow, and whether any of the outflow is emitted by extended regions outside the compact nuclear source, we make two extreme assumptions. First, we assume that all the narrow [OIII] emission in the  spectroscopic aperture is emitted by the compact nuclear source. In this case, based on a comparison between our imaging and spectroscopic results within the spectroscopic aperture, we deduce that 80\% of the outflowing [OIII] (broad and intermediate components) must be emitted by the compact structure. Second, we assume that all the narrow [OIII] is emitted outside the compact nucleus. In this case, 85\% of the outflowing [OIII] must be emitted by the compact structure. The second assumption is more plausible given the likelihood that the degree of kinematic disturbance will increase towards the nucleus. Therefore we deduce that the the overwhelming majority, of the nuclear outflow is emitted by the compact source, and adopt the instrumentally corrected HWHM of the compact source (69$\pm$1 pc; see Table \ref{table:2dgauss}) as the radius of the outflow.

Note that, by adopting the HWHM as the radius of the [OIII] outflow, we are assuming that it is symmetric about the AGN nucleus. However, this is not necessarily the case: there is evidence in F13451+1232 that the cooler phases of the outflow are more  extended on the south side  of the nucleus, since \citet{morganti13} show that the neutral HI and molecular CO outflows are concentrated at the extreme of the southern radio jet ($\sim$130~pc to the south of the nucleus), but find no evidence for a molecular or HI outflow along the shorter radio jet to the north. If the molecular, neutral  and ionized outflows are  co-spatial in F13451+1232 -- as they appear to be in at least one AGN with multi-phase outflows that have been mapped in detail \citep[IC5063:][]{morganti07,tadhunter14,morganti15} -- then the ionized gas outflow might also be be asymmetric.  Indeed, such asymmetry would be consistent with the fact that the FWHM of the compact [OIII] structure visible in our {\it HST} images (138~pc) is similar to the radial extent of the molecular outflow to the south of the nucleus. Therefore, the [OIII] outflow could be more extended than the 69~pc indicated by the HWHM.

\subsubsection{IRAS F14394+5332} 

\begin{figure}
\centering
\includegraphics[width=0.5\textwidth]{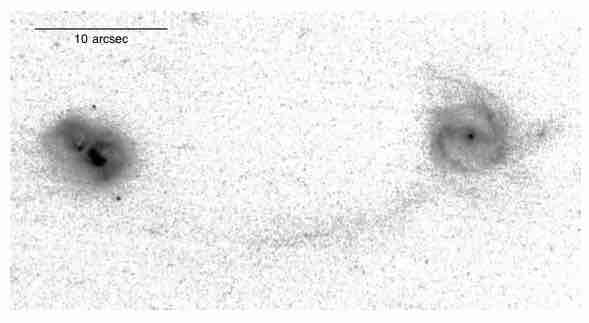}
\caption{A 43$\times$22 arcsec FoV including the two merging galaxies of the
  system IRAS F14394+5332.}
\label{14394all}
\end{figure}

Figure \ref{14394all} shows a wide field-of-view image of
the continuum emission from the two interacting galaxies that form this system. The nuclear separation between the
two galaxies is $\sim$28 arcsec ($\sim$56\,kpc), and they are connected by a bridge
of faint, diffuse emission. The western component shows a spiral
morphology, and no [OIII] emission is detected from this source. Therefore, it
is not shown in Figure \ref{HST_frames}.

A priori, the morphology of the continuum emission from the eastern component in our {\it HST} images is
consistent with that described in \cite{kim02} in which the eastern
component is a double nucleus galaxy in itself, with the brighter nucleus (labelled E) separated
from a fainter nucleus (labelled EE) 2.0 arcsec (3.7\,kpc) to the NE along PA67. 
The \citet{kim02} results were based in near-IR K-band images. 
However, the double nucleus nature of the eastern component is not so
clear in our higher resolution {\it HST} optical images, which present a
complex morphology that includes condensations, tails and dust lanes; it is likely 
that the two brightest condensations in this image --- which appear to be divided by a 
dust lane aligned along PA142 --- form the E nucleus described in
the \citet{kim02} paper.

Spectroscopically, the E nucleus is classified as a Seyfert 2 galaxy at optical
wavelengths \citep{veilleux99,yuan10}. RZ13 carried out a
spectroscopic study of the nuclear outflows in this nucleus and found that it
shows the most spectacular [OIII] emission line kinematics of all the
objects considered in their study (see their Figure 1). In order to fit the [OIII] emission line profile, three narrow (FWHM $<$ 450 km/s) and two broad (FWHM $\sim$ 1300 km/s) components with extreme velocity shifts of up to $\sim$1500 km/s were required.
If we assume that the reddest of the three narrow components represents the rest frame of the host galaxy -- this is supported
by observations of the extended emission lines on either side of the nucleus  -- and that all the other (blueshifted) components represent the outflow, then the outflow contributes 87\% of the total [OIII] flux.

Our {\it HST} [OIII] image of the system shows an irregular structure without a clear compact nucleus. The highest surface brightness
part of this structure is centred on the more northerly of the two
continuum condensations associated with the E nucleus, and to the south the structure is cut through by the dust lane that divides
the two main continuum condensations. No [OIII] emission is detected in the vicinity of the EE nucleus
of \citet{kim02}. 

At the resolution of our {\it HST} images, the true position of the AGN nucleus is uncertain. If the AGN is associated
with the peak of the [OIII] emission on the NE side of the dust lane, the [OIII] emission extends to a maximum radius of
1.0 arcsec (1.9\,kpc) and the flux weighted mean radius of the outflow is 0.36 arcsec (0.65\,kpc). On the other hand, if the
AGN is located in the dust lane, between the two continuum condensations, the [OIII] emission has a maximum radius of 1.2 arcsec (2.2\,kpc) and a flux weighted mean radius of 0.45 arcsec (0.84\,kpc). 




\subsubsection{IRAS F15130-1958} 

The {\it HST} continuum image of IRAS F15130-1958 shows
a complex nuclear morphology, with various dust features, as well as a tidal tail
that extends $\sim$8 arcsec ($\sim$16\,kpc) to the SE of the galaxy. Only one prominent,
nuclear condensation of continuum emission is observed in the image.

Our [OIII] image of this source is dominated by a
compact structure of ionized gas emission centred on the main nuclear continuum concentration. There is also some
faint, diffuse emission to the south of the nucleus that extends to a maximum radius of  0.40 arsec (0.77\,kpc). 

The spectroscopic aperture of RZ13 (1.5$\times$2.4 arcsec aperture along PA358) encompass the entire compact [OIII] structure and the diffuse emission. RZ13 modelled the [OIII] emission line profiles using a combination of two intermediate ($FWHM  = 545\pm85$ and $FWHM = 700\pm222$ km s$^{-1}$) components, and one broad (FWHM: $1630\pm42$ km s$^{-1}$) component. Taking the narrowest, most redward intermediate component to represent the galaxy rest frame, the bluer intermediate and broad components are shifted by  -350$\pm$280 and -725$\pm$131 km s$^{-1}$. However, using higher resolution Xshooter data we have recently measured the rest frame of the host galaxy more accurately using the high order Balmer and near-IR CaII triplet stellar absorption lines \citep{rose17}, and obtain an adequate fit to the [OIII] emission lines using a combination of two Gaussians of width $828\pm38$~km s$^{-1}$ and $1250\pm190$~km s$^{-1}$ (FWHM) shifted by $-465\pm15$ and $-1095\pm70$~km s$^{-1}$ respectively relative to the rest frame defined by the stellar absorption lines. Therefore, in this case the entire [OIII] emission feature is blueshifted and associated with the warm outflow; there is no evidence for a narrow component at rest-frame wavelength defined by the stellar absorption features.

Based on analysis of the narrow-band [OIII] image using circular aperture photometry, we find that the total flux in the compact nuclear source within an aperture of 3.5 pixel (0.175 arcsec) radius is $1.4\times10^{-14}$ erg s$^{-1}$ cm$^{-2}$. Following aperture correction by a factor of 1.22 \citep{sirianni05}, the total [OIII] flux associated with the compact nuclear source is $1.7\times10^{-14}$~erg s$^{-1}$ cm$^{-2}$. This is similar to the total [OIII] flux measured within a circular aperture that has the same radius (1.3 arcsec, 26 pixels) as that of the 
spectroscopic extraction aperture: $1.9\times10^{-14}$~erg s$^{-1}$ cm$^{-2}$. Therefore, the compact nuclear [OIII] source contributes 90\% of the total flux in the nuclear aperture. Clearly the flux contribution of the diffuse structures to the south of the compact nucleus is relatively minor.

Given that our spectra show that the entire [OIII] emission line profile is associated with the outflow, and our {\it HST} images demonstrate that  
the overwhelming majority of the [OIII] flux is emitted by the compact structure, we again adopt the instrumentally corrected HWHM of the 
compact structure (66$^{+8}_{-10}$ pc; see Table \ref{table:2dgauss}) as the radius of the outflow.

\subsubsection{IRAS F16156+0146} 

The system is a merger between two galaxies with
a nuclear separation of $\sim$3.5 arcsec ($\sim$7.9\,kpc) in the N-S direction. The {\it HST}
continuum image of the source shows that the morphology of each of the two merging
galaxies consist of a central condensation plus an irregular halo of diffuse
emission. No prominent dust lanes and/or additional tidal features are visible
in our continuum image of the source. However, the [OIII] image of the galaxy
shows a series of clumps or condensations embedded in more diffuse emission. Relative to the NW nucleus, 
these clumps extend $\sim$0.8 arcsec
($\sim$1.8 kpc) in the NW direction, and 1.9 arcsec ($\sim$4.3\,kpc) in the SE direction along PA$\sim$130. This [OIII] structure extends almost the entire
distance between the two merging galaxies, and its close alignment with the axis joining the nuclei suggests that it is a tidal feature. 

The brightest [OIII] clump coincides with
the continuum nucleus of the NW source, which is classified as Sy2 at optical wavelengths
\citep{veilleux99,yuan10}. RZ13 found evidence for an AGN-induced outflow in their 
1.5$\times$2.2 nuclear aperture centred on the NW nucleus. Since the longest axis
of this aperture is aligned along PA133, it contains much of the
[OIII] emission visible in the {\it HST} image. The nuclear [OIII] profile is well-fitted with three
Gaussian components: one narrow ($FWHM < 200$~km s$^{-1}$), one intermediate ($FWHM = 804\pm26$~km s$^{-1}$) and one broad ($FWHM = 1535\pm63$~km s$^{-1}$), with the intermediate and broad components shifted from the narrow component by $-186\pm10$~km s$^{-1}$ and $-374\pm26$~km s$^{-1}$ respectively. Taking the  intermediate and broad components to represent the outflow, these components together contribute 69\% of the total [OIII] flux in the spectroscopic
aperture.

The [OIII] flux measured within a 3.5 pixel (0.175 arcsec) radius aperture centred on the NW compact nucleus in our {\it HST} image is $1.3\times10^{-14}$~erg s$^{-1}$ cm$^{-2}$. Assuming an aperture correction factor of 1.22 \citep{sirianni05}, this component has a total [OIII] flux of $1.6\times10^{-14}$~erg s$^{-1}$ cm$^{-2}$. In comparison, the total [OIII] flux measured within a 22 pixel (1.1 arcsec) radius aperture, corresponding to the extent of the spectroscopic aperture is $2.9\times10^{-14}$~erg s$^{-1}$ cm$^{-2}$. Therefore the compact [OIII] nucleus in this source contributes 54\% of the [OIII] flux within the spectroscopic aperture. Following the method used for PKS1345+12, based on the comparison between the spectroscopic and the imaging data, we deduce that 33 -- 78\% of the flux in the outflow is emitted by the compact nucleus. However, although a large fraction of the [OIII] flux could be emitted by the compact nucleus, this is obviously not such a clear-cut case as those of F13451+1232 and F15130-1950 described above. The maximum possible radius of the [OIII] outflow in this case corresponds to the maximum extent of the [OIII] structures visible in the {\it HST} image (4.3\,kpc); however, we have no spectroscopic evidence that [OIII]-emitting regions outside our spectroscopic aperture (radius 2.5~kpc) are outflowing. The flux weighted mean radius measured in a 5\,kpc diameter aperture centred
on the compact nucleus is 0.849$\pm$0.004\,kpc.

\subsubsection{IRAS F17044+6720} 

The {\it HST} continuum image of this source
shows a highly disturbed morphology. Two condensations of enhanced continuum
emission, separated by $\sim$0.39 arcsec ($\sim$0.90\,kpc), are visible in the
central region of the source, which might correspond to the two nuclei of the
merging galaxies, or a single nucleus bisected by a dust lane. 
A series of dust features is also observed through almost the
entire extent of the system, although they are more prominent as they cross
the ``circular'' halo of diffuse continuum emission in the nuclear region. In
addition, a tidal tail of length $\sim$6 arcsec ($\sim$14\,kpc) emerges from the nuclear
region to the north, and then bends to the west. Several clumps are visible within this
tail structure. 


In the case of IRAS F17044+6720, the {\it HST} [OIII] image is truly
spectacular. Regions of enhanced ionized gas emission are observed in the
nuclear region, and then extended $\sim$4.0arcsec ($\sim$9.2\,kpc) in the direction of the tidal
tail, along PA$\sim$155; this association with the continuum tidal tail suggests that the general morphology of emission line gas is tidal in origin, and is not shaped by other mechanisms such as outflows origination from the nucleus, or the anisotropy of the AGN radiation field. Three bright,[OIII]-emitting clumps are visible in the tail structure (see Figure \ref{HST_frames}) that we label clumps A, B, C  at distances 1.8, 2.8, 3.7  arcseconds ($\sim$4.2, 6.4, 8.5\,kpc) from the nucleus respectively.  Note that the brightest [OIII] clump in the nuclear regions is centred on the more northerly of the two continuum condensations visible in the continuum image. We take the latter to represent the AGN nucleus.

The long-slit WHT/ISIS observations discussed in RZ13 were taken with the slit aligned along PA160 --- covering the nucleus as well as the emission line clumps to the north of the nucleus. RZ13 extracted 1.5$\times$2.2 arcsec nuclear aperture centred on the double continuum nucleus to the south. This 
aperture includes the brightest condensations in the nucleus, as well as the more diffuse structure to the N and W, but avoids the bright clumps (A, B and C) to the N of the nucleus. RZ13 found that the [OIII] emission-line profile from this nuclear aperture could be modelled using a combination of narrow ($FWHM = 290\pm20$~km s$^{-1}$) and broad ($FWHM = 1765\pm103$~km s$^{-1}$) Gaussian components, with the latter shifted by $-553\pm65$~km s$^{-1}$ relative to the narrow component and contributing 36\% of the total
[OIII] flux in nuclear aperture. We identify the broad, blue-shifted component as the outflow in this object.

We have extracted further apertures from our PA160 long-slit spectrum in order to examine whether
they show any evidence for outflows. The apertures were centred on clumps A, B and C. Following subtraction of the stellar continuum using the {\sc STARLIGHT} code \citep{mateus06},
the Gaussian profiles were fitted to the emission lines and emission line blends using the {\sc STARLINK} {\sc dipso} package. 
We found that a single, narrow Gaussian was sufficient to provide a good fit to the profiles of all the emission lines with the exception of the [OIII] emission in clump B. In the latter case, the [OIII] emission lines required a second broader, blue-shifted component in order to fit the profiles. The broader component has a velocity shift of -144$\pm$27 relative to the narrow component and a width $FWHM=485\pm17$ km s$^{-1}$. However, the velocity shift and line width measured for the broader component detected in clump B are much smaller than those measured for the [OIII] emission in the nucleus ($\Delta$v=-553$\pm$63 km/s and $FWHM=1765\pm$103 km s$^{-1}$; RZ13). Therefore, they do not provide clear evidence for an extended
outflow.

\begin{figure}
\centering
\includegraphics[scale=0.3, trim=0.1cm 4.6cm 0.1cm 3.6cm]{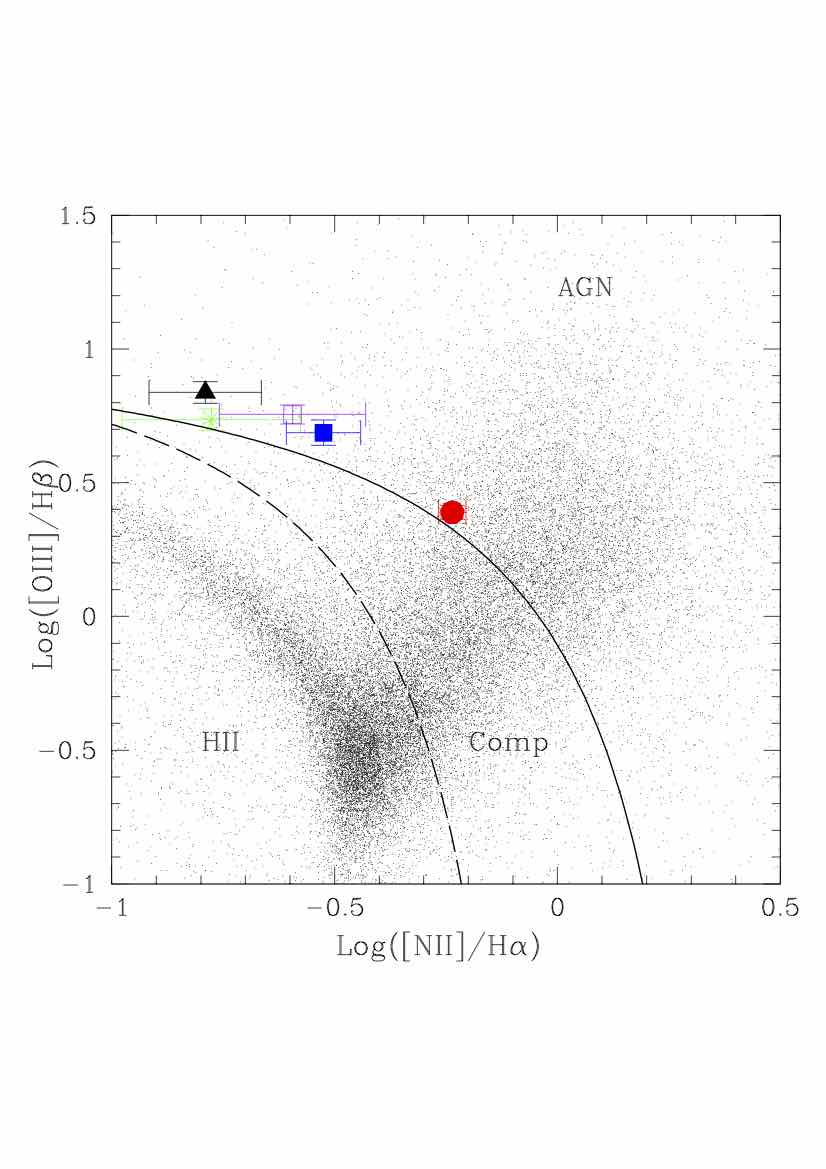}
\caption{Diagnostic plot of Log$_{10}$([OIII]/H$\beta$) vs Log$_{10}$([NII]/H$\alpha$). AGN are defined to lie above the solid line, HII-region like galaxies below the dashed line, and composite galaxies between these boundaries, following the
definitions of \citet{kewley06}. The red circle represents the ratio for the nuclear emission, the blue square represents clump A, the green asterisk represents clump B, the black triangle represents clump C, and the open purple square represents the total clump emission. The small points represent objects detected by the SDSS. 
}
\label{fig:bpt}
\end{figure}

Both the starburst and the AGN activity could potentially ionize the extended clumps in the ENLR of this object. Therefore, to investigate the ionisation mechanism we have plotted the emission line ratios on Baldwin, Phillips and Terlevich (BPT) diagnostic diagrams \citep{bpt}. Figure \ref{fig:bpt} shows the results for the Log$_{10}$([OIII]/H$\beta$) vs Log$_{10}$([NII]/H$\alpha$) diagram. The line ratios measured for all the apertures fall in the AGN part of the diagram, suggesting that AGN photoionisation
dominates. However, the points for the extended apertures fall in part of the diagnostic diagram that is relatively unpopulated
for AGN nuclei. As shown by the photoionisation modelling of \citet{groves06}, this is consistent with relatively low abundances 
in the ENLR -- perhaps as 
low as 0.5Z$_{\odot}$. Given the well-known mass-metallicity relationship for galaxies \citep{tremonti04} and the presence of substantial radial metallicity gradients in spiral galaxies such as the Milky Way, this could imply that the ENLR material has been 
accreted from the central regions of low mass precursor galaxy, or from the outer disk of a more massive precursor.

We have used the {\it HST} [OIII] image to investigate how much of the outflow in the nuclear regions is emitted by
the brightest clump centred on the nucleus. Within a 3.5 pixel (0.175 arcsec) radius aperture centred on this
clump we measure an [OIII] flux of $1.4\times10^{-15}$~erg s$^{-1}$ cm$^{-2}$, corresponding to a total
flux emitted by this component of $1.7\times10^{-15}$~erg s$^{-1}$ cm$^{-2}$ assuming an aperture correction of 1.22
\citep{sirianni05}. For comparison, the total [OIII] flux in a 22 pixel (1.1 arcsec) aperture 
centred on the nucleus is $6.8\times10^{-15}$~erg s$^{-1}$ cm$^{-2}$. Therefore the compact nucleus contributes 25\% of the [OIII] flux in the nuclear aperture. Using the same arguments as used above for F1345+12 and F15130-1958
we deduce that 0 -- 69\% of the outflow could be emitted by the compact nucleus. Note, however, that, due to the complexity
of the near-nuclear structures in this case, it was not possible to fit a 2D Gaussian in order to determine the
radius of the compact nucleus.
The flux-weighted mean
radius of the [OIII] emission within a circular aperture of diameter 5\,kpc centred on the nucleus is 1.2$\pm$0.01\,kpc.

\subsubsection{IRAS F17179+5444} 

The {\it HST} continuum image of F17179+5444
shows a complex morphology. Two condensations, separated by $\sim$0.47
arcsec ($\sim$1.16\,kpc) and connected by a bridge of enhanced continuum emission,
are observed in the nuclear region of the galaxy. These condensations are
surrounded by a semi-circular halo of diffuse emission, where a series of
dust features is clearly visible. In addition, there is a faint tail of
continuum emission emerging from the south of the nuclear region and extended
$\sim$6 arcsec ($\sim$15\,kpc) towards the SE of the galaxy.

In contrast to the continuum structure, the [OIII] emission from the galaxy shows
four clumps of ionized gas emission. The compact northern clump is substantially brighter
than the other three, and coincides with the location of the brighter, southern condensation
in the nuclear continuum emission. We assume that the latter corresponds to the AGN nucleus.   
A second region of enhanced [OIII] emission emerges $\sim$0.4 arcsec
to the south of the northern clump, and two additional condensations $\sim$0.8
arcsec ($\sim$2.0\,kpc) to the SE and SW of the northern clump; the maximum extent of the emission
line structures is 1.1 arcsec (2.6\,kpc).

The nuclear spectroscopic aperture of RZ13 (1.5$\times$2.0 arcsec along PA122) encompasses the compact [OIII] structure and the extended clumps to the south, south east. Within this spectroscopic aperture, the [OIII] emission line profile  can be modelled using a combination of a narrow component ($FWHM = 358\pm75$ km~s$^{-1}$), an intermediate component ($FWHM=515\pm33$ km~s$^{-1}$) and  broad component($FWHM=1562\pm43$ km~s$^{-1}$), with the intermediate and broad components shifted by $+123\pm34$ and $-242\pm61$ km~s$^{-1}$ respectively relative to the rest frame defined by the narrow component.
If the broad and intermediate components are together taken to represent the outflow, the  outflow contributes 88\% of the total [OIII] flux in the nuclear aperture (RZ13).


Performing aperture photometry on our {\it HST} [OIII] image,  we find that the flux of compact nuclear source within an aperture of 3.5 pixel (0.175 arcsec) radius is  $9.3\times10^{-15}$ erg s$^{-1}$ cm$^{-2}$. Assuming an  aperture correction  factor of 1.22 \citep{sirianni05}, the total [OIII] flux associated with the compact nuclear source is estimated to be $1.1\times10^{-15}$ erg s$^{-1}$ cm$^{-2}$. This comprises of 85\% of the total [OIII] flux measured in a circular aperture with a 1.0 arcsec radius equivalent to that of the spectroscopic aperture.

Adopting a similar approach to that used for PKS1345+12 and F16156-0146 above, from the comparison between the results from the {\it HST} imaging and ground-based spectroscopy we deduce that between 83 and 97\% of the flux of the nuclear outflow components in F17179+5444 must be emitted by the compact nuclear structure. Therefore we adopt the instrumentally corrected HWHM of this structure (112$\pm$7\,pc; see Table \ref{table:2dgauss}) as an estimate of the radius of the outflow region.

\section{Discussion and conclusions}

Based on the results of our {\it HST} imaging programme, the warm gas outflows associated with the AGN in our complete sample of ULIRGs are mostly compact: in three objects (PKS1345+12, F15130--1958 and F17179+5444) they are barely resolved with {\it HST} and have radii in the range 0.065 -- 0.12\,kpc; and in a further five objects (F13428+5608, F13443+0802, F14394+533, F16156+0146 and F17044+6720) the outflows are spatially resolved but with maximum radii in the range 1.9 -- 5\,kpc, and flux-weighted mean radii in the range 0.65 -- 1.8\,kpc. These results are consistent with the measurements of, or upper limits on, the outflow radii derived from our long-slit spectra in papers I \& III for the wider QUADROS sample \citep{rose17,spence18}. Therefore we do not find evidence in our data for the galaxy-wide outflows predicted by some models of AGN outflows. This point is dramatically illustrated by the right-hand panels in Figure 1. These show that the [OIII] structures -- not all of which are necessarily outflowing \citep{rz14,spence18} -- cover only a small fraction of the extended starlight structures of the host galaxies in most cases.

It is also interesting that the radial extents that we measure for the warm ionized outflows in our sample objects are similar to those typically estimated for the cool molecular outflows in nearby ULIRGs using CO emission lines and OH absorption lines:   0.05 -- 1.23\,kpc \citep[][but see Veilleux et al. 2017]{cicone14,gonz17}. Thus it is possible that warm ionized and cool molecular outflow regions are co-spatial, although high resolution molecular line observations will be required to confirm this, by investigating the spatial relationship between the different phases of the outflows for the individual objects in our sample.

These results also inform the ongoing debate about the true extent of the warm outflows associated with AGN. A number of integral field studies of AGN at both high and low redshifts have found evidence for outflows that extend on scales
of 2 -- 15\,kpc \citep{harrison12,liu13,harrison14}. The integral field study of \citet{harrison12}, which finds typical outflow radii for a sample of high-redshift ($1.4 < z < 3.0$) ULIRGs in the range 2 -- 10\,kpc, is particularly relevant to the current study. In contrast, other studies that are based both on integral field and long-slit data do not find evidence for AGN-driven outflows extended on scales $>$5\,kpc in most cases. For example, in their re-analysis of the integral field data presented in \citet{liu13}, \citet{husemann16} stress the importance of correcting for the beam smearing effects caused by atmospheric seeing when measuring the radii of the outflow regions, and find that
the outflow regions could be more compact than has previously been claimed. Moreover, some of the integral field studies have  deliberately targetted samples  of AGN that were already known to have extreme nuclear outflows based on previous spectra, perhaps biasing the studies towards larger outflow radii. 

We emphasise that the sample used for the current study, while relatively small, is complete and representative of local ULIRGs with optical AGN nuclei. The hydrodynamic simulations of galaxy mergers predict that AGN-driven outflows should be particularly important in just such objects. The fact that we find that the outflows are relatively compact in these extreme objects further supports the idea that the effects of the warm, AGN-driven outflows associated with powerful AGN are localised to the near-nuclear regions.

Finally, we note that our results on the radial extents of the warm outflows in ULIRGs relate to the high-surface-brightness [OIII] structures that dominate the emission-line fluxes in both our HST images and ground-based spectra. However, we cannot rule out the idea that there exists a more diffuse, low-surface brightness, and low density component that represents a significant proportion of the total outflow mass but contributes relatively little to total emission-line flux \citep[see discussion in][]{spence18}. Indeed, even in the three cases in which a compact nuclear source dominates the {\it HST} [OIII] images, comparison with our optical spectra demonstrates that up to 10 -- 20\% of the [OIII] flux of the outflow could be emitted by a spatially diffuse component.

\section*{Acknowledgements}
MR \& CT acknowledge support from STFC, and CRA acknowledges the Ram\'{o}n y Cajal Program of the Spanish Ministry of Economy and Competitiveness through project RYC-2014-15779 and the Spanish Plan Nacional de Astronom\'{i}a y Astroﬁs\'{i}ca under grant AYA2016-76682-C3-2-P. Based on observations taken with the NASA/ESA Hubble Space Telescope, obtained at the Space Telescope Science Institute (STScI), which is operated by AURA, Inc. for NASA under contract NAS5-26555. The STARLIGHT project is supported by the Brazilian agencies CNPq, CAPES and FAPESP and by the France-Brazil CAPES/Cofecub program. The authors acknowledge the data analysis facilities provided by the Starlink Project, which was run by CCLRC on behalf of PPARC. This research has made use of the NASA/IPAC Extragalactic Database (NED) which is operated by the Jet Propulsion Laboratory, California Institute of Technology, under contract with the National Aeronautics and Space Administration.

\end{document}